\documentclass[%
 reprint,
 amsmath,amssymb,
 aps,
 prb,
]{revtex4-1}

\usepackage{graphicx}% Include figure files
\usepackage{dcolumn}% Align table columns on decimal point
\usepackage{bm}% bold math
\usepackage{hyperref}% add hypertext capabilities
\begin{document}

\preprint{APS/123-QED}

\title{Exploring Dynamical Phase Transitions in the XY Chain through Linear Quench: Early and Long-term Perspectives}% Force line breaks with \\

\author{Kaiyuan Cao}
\email{191001004@njnu.edu.cn}
\affiliation{Zhejiang Lab, Hangzhou 311100, People's Republic of China}

\author{Peiqing Tong}
\email{pqtong@njnu.edu.cn}
\affiliation{Department of Physics and Institute of Theoretical Physics, Nanjing Normal University, Nanjing 210023, People's Republic of China}
\affiliation{Jiangsu Key Laboratory for Numerical Simulation of Large Scale Complex Systems, Nanjing Normal University, Nanjing 210023, People's Republic of China}

\date{\today}% It is always \today, today,

\begin{abstract}
  We investigate the nonequilibrium dynamics induced by a finite-time linear quench in the XY chain. Initially, we examine the dynamical quantum phase transition, characterized by the nonanalytic behavior of the Loschmidt amplitude. We find distinct behaviors of DQPTs during and following the ramp. Following the ramp, the ramp crossing the critical point $h_{c}$ is the sufficient condition for the occurrence of DQPT, but it is not during the ramp. Through AIA approximation analysis, we establish that adequate distancing from the critical point is crucial for DQPT manifestation during the ramp, elucidating the absence of DQPT as the ramp gets faster. Additionally, we explore another type of dynamical phase transition, describing the long-term relaxation behavior of the order parameter. Our finding indicates that the asymptotic behavior of the time-dependent part induced by the linear quench is equivalent to that following a sudden quench, i.e., time-dependent part exhibits power-law decays of $\sim t^{-3/2}$ and $\sim t^{-1/2}$ for the ramp to the commensurate and incommensurate phases, respectively. Moreover, we also delve into the steady part, which showcases nonanalytic singularities at the critical point.
\end{abstract}

\maketitle

%\tableofcontents

\section{Introduction}

Phase transitions are intrinsic phenomena in physics, characterized by sudden changes in a system's properties as external parameters vary \cite{Sondhi1999rmp, Sachdev2011}. The investigation of phase transitions has provided valuable insights into the behavior of a diverse range of physical systems. Recently, there has been a surge of interest in non-equilibrium phase transitions, which occur in systems undergoing dynamics away from equilibrium \cite{Calabrese2006prl, Mondal2010, Dziarmaga2010advp, Polkovnikov2011rmp}. These transitions deviate from traditional equilibrium phase transitions, making them an intriguing area of research. Within the realm of dynamical phase transitions, two notable phenomena have attracted significant attention: dynamical quantum phase transitions (DQPTs) \cite{Heyl2013prl, Zvyagin2016ltp, Heyl2018rpp, Heyl2019el} and dynamical phases characterized by the relaxation behavior of order parameters \cite{Prosen2008prl, Eckstein2009prl, Barmettler2009prl, Barmettler2010njp, Dziarmaga2010advp, Marino2012prb}. The former entails the non-analytical evolution of the Loschmidt amplitude at early times, offering a unique perspective on the interplay between quantum dynamics and critical phenomena. The latter delineates dynamical phases based on the distinct relaxation behavior of order parameters towards their equilibrium states following prolonged evolution. Exploring these two types of dynamical phase transitions, from both early-time and long-time perspectives, is crucial for unraveling the intricate behavior of physical systems and elucidating the fundamental principles governing their dynamics.

In the study of nonequilibrium dynamics, due to the advancement of cold atomic experiments \cite{Greiner2002nature, Kinoshita2006nature, Bloch2008rmp, Georgescu2014rmp}, isolated quantum systems subjected to a quantum quench \cite{Calabrese2006prl, Calabrese2007jsmte, Iucci2009pra, Manmana2009prb, Brandino2012prb, Rigol2014prl, Collura2014prb} and periodic driving processes \cite{Lindner2011nphys, Sacha2018rpp, Kurzyna2020prb} have received more attention. From the early-time perspective, many studies have explored the basic features of DQPTs through designed quantum quenches \cite{Heyl2013prl, Andraschko2014prb, Hickey2014prb, Vajna2014prb, Schmitt2015prb, Vajna2015prb, Divakaran2016pre, Bhattacharya2017prb, Kosior2018pra, Lang2018prb, Lahiri2019prb, Liu2019prb, Cao2020prb, Cao2023prb, Kuliashov2023prb, Sacramento2024prb} %and periodic driving \cite{Yang2019prb, Jafari2022prb, Naji2022prb, Naji2022pra} 
in various quantum systems, in which the system's parameters are changed suddenly. In contrast, the slow quench \cite{Dziarmaga2005prl, Dziarmaga2010advp, Grandi2010prb, Weiss2018pra, Balducci2023prl} as a natural extension of the sudden quench has received limited attention in the existing literature. In Refs.~\onlinecite{Divakaran2016pre, Sharma2016prb, zamani2310}, the authors treat the slow quench as a way to prepare the initial state and solely study the DQPT following the ramp. Ref.~\onlinecite{Tatjana2016SciPost} discussed the behavior of the Loschmidt amplitude during the slow quench in the transverse Ising chain. They find that the DQPT only occurs when the ramp is slow, i.e., the quench time $\tau_{Q}$ is large. However, the absence of DQPT for the rapid quench remains unclear. This discrepancy is of interest, as rapid quench is conceptually closer to a sudden quench than a slow quench.

On the other hand, from the long-term perspective, a new type of dynamical phase transition (DPT) has been proposed to characterize the dynamical phase with the asymptotic scaling behavior of order parameters approaching their steady values \cite{Sen2016prb, Sourav2018jpa, Sarkar2020prb, Aditya2022prb, Makki2022prb, Zou2023prb, Cao2024prb109}. By taking the fermion two-point correlation function $C_{mn}(t)=\langle c_{m}^{\dag}(t)c_{n}(t)\rangle$ as an example, the deviation $\delta C_{mn}(t) = C_{mn}(t)-C_{mn}(\infty)\sim t^{-\mu}$ of $C_{mn}(t)$ from its steady state value $C_{mn}(\infty)$ transients with time as a power-law decay $\sim t^{-\mu}$. In the XY chain, the scaling exponent of $\delta C_{mn}(t)$ after a sudden quench is determined by whether the post-quench Hamiltonian is in the commensurate phase ($\sim t^{-3/2}$) or the incommensurate phase ($\sim t^{-1/2}$) \cite{Makki2022prb}. We note that Ref.~\onlinecite{Tatjana2016SciPost} studied the relaxation behavior of the similar transverse magnetization and correlation functions in transverse Ising chain and found the power-law decay $\sim t^{-3/2}$. This inspires us to speculate that the DPT of the XY chain in the sudden quench can be extended to slow quench processes.

In this paper, we investigate the dynamics of the linear quench in the XY chain from both the early-time and long-time perspectives. We consider the finite-time linear quench different from that in the previous literature \cite{Tatjana2016SciPost}. While the previous study \cite{Tatjana2016SciPost} limited the annealing time to a fixed quench time $\tau_{Q}$, our model incorporates a ramp duration dependent on the final external field $h_{f}$. In the study of DQPTs, by calculating the rate function and Pancharatnam geometric phase (PGP), we identify distinct behaviors of DQPTs during and following the ramp, offering insights into the absence of DQPT with faster linear quenches \cite{Tatjana2016SciPost}. In the study of the second type of DPT, we derive the expression of the fermion two-point correlation function, and thus establish that the long-time asymptotic behavior of the time-dependent part is solely determined by whether the final parameter lies in the commensurate or incommensurate phase. We successfully elucidate the dynamic behavior of linear quenches from both early- and long-time perspectives using the AIA approximation.

The paper is organized as follows: In Sec.~\ref{sec:model}, we introduce the XY chain and derive the time-dependent state of system during and following the linear quench. In Sec.~\ref{sec:Le}, we investigate the behavior of DQPT during the linear quench. In Sec.~\ref{sec:order-parameter}, we study the relaxation behavior of the fermion two-point correlation $C_{mn}(t)$ following the linear quench. We conclude in Sec.~\ref{sec:conclusion}

\section{model}
\label{sec:model}

The Hamiltonian for the XY chain in the transverse field can be expressed by
\begin{equation}\label{eq:XY-Hamiltonian}
  H  = -\frac{1}{2}\sum_{n=1}^{N}(\frac{1+\gamma}{2}\sigma_{n}^{x}\sigma_{n+1}^{x}+\frac{1-\gamma}{2}\sigma_{n}^{y}\sigma_{n+1}^{y}+h\sigma_{n}^{z}), 
\end{equation}
where $\sigma_{n}^{x,y,z}$ are the Pauli operators defined on the lattice site $n$, $\gamma$ represents the anisotropic parameter, and $h$ denotes the external magnetic field. Under the limits $\gamma \rightarrow 0$ and $\gamma \rightarrow 1$, the system reduces the isotropic XY chain and the transverse-field Ising model, respectively. The Hamiltonian (\ref{eq:XY-Hamiltonian}) can be mapped to a quadratic fermion model via the Jordan-Wigner transformation
\begin{equation}
    H = -\frac{1}{2}\sum_{n=1}^{N}[(c_{n}^{\dag}c_{n+1} + \gamma c_{n}^{\dag}c_{n+1}^{\dag} + h c_{n}^{\dag}c_{n}) + h.c.].
\end{equation}
With the periodic boundary condition, we obtain the Hamiltonian in the momentum space after using the Fourier transformation as
\begin{equation*}
    \begin{split}
        H & = \sum_{k>0}H_{k} = \sum_{k>0}\Psi_{k}\mathbb{H}_{k}\Psi_{k}^{\dag} \\
          & = \sum_{k>0}\left(
                          \begin{array}{cc}
                            c_{k}^{\dag} & c_{-k} \\
                          \end{array}
                        \right)
                        \left(
                          \begin{array}{cc}
                            -h-\cos{k} & i\gamma\sin{k} \\
                            -i\gamma\sin{k} & h+\cos{k} \\
                          \end{array}
                        \right)
                        \left(
                          \begin{array}{c}
                            c_{k} \\
                            c_{-k}^{\dag} \\
                          \end{array}
                        \right).
    \end{split}
\end{equation*}
By applying the Bogoliubov transformation, $\eta_{k} = u_{k}c_{k}+v_{k}c_{-k}^{\dag}$, we can immediately arrive at the diagonalized form of the Hamiltonian 
\begin{equation}
    H = \sum_{k}\varepsilon_{k}(\eta_{k}^{\dag}\eta_{k}-\frac{1}{2}),
\end{equation}
where the quasiparticle excitation spectrum is $\varepsilon_{k} = \sqrt{(h+\cos{k})^{2}+\gamma^{2}\sin^{2}{k}}$. The ground state can be expressed as the BCS-like form
\begin{equation}\label{eq:ground-state}
    |\mathrm{GS}\rangle = \prod_{k>0}(u_{k}+v_{k}c_{k}^{\dag}c_{-k}^{\dag})|0\rangle,
\end{equation}
where 
\begin{eqnarray}
    u_{k} &=& \cos{\theta_{k}} = \frac{\varepsilon_{k}-h-\cos{k}}{\sqrt{2\varepsilon_{k}(\varepsilon_{k}-h-\cos{k})}}, \\
    v_{k} &=& i\sin{\theta_{k}} = \frac{i\gamma\sin{k}}{\sqrt{2\varepsilon_{k}(\varepsilon_{k}-h-\cos{k})}}
\end{eqnarray}
are given by defining the Bogoliubov angle $\theta_{k}$, and $|0\rangle$ is the fermion vacuum state satisfying $c_{k}|0\rangle=0$ and $c_{-k}|0\rangle$. For simplicity, we can also express $c_{k}^{\dag}c_{-k}^{\dag}$ to be $|1\rangle$, which satisfies $c_{k}^{\dag}|1\rangle = 0$ and $c_{-k}^{\dag}|1\rangle = 0$.

The previous literature has verified that the XY model undergoes the second-order phase transition when its parameter passes through the quantum critical point (QCP) \cite{Lieb1961407}. In the second-order phase transition, the second-order phase transition can be characterized by an energy gap between the ground state and the first excited state vanishes at the QCP, i.e. $\varepsilon_{k}=0$. In our model, it is easy to obtain that the system has QCPs at $h_{c}=\pm1$, and $\gamma_{c}=0$ for $|h|\leq1$. Moreover, we need to mention that there is another type of transition called the commensurate-incommensurate transition in the XY model \cite{Barouch1971pra, Bunder1999prb}. In the commensurate phase, spin correlation functions of the XY chain have non-oscillatory asymptotic behavior, while in the incommensurate phase, spin correlation functions show the oscillatory behavior \cite{Barouch1971pra}. Notably, the boundary of the commensurate-incommensurate transition can be identified by where the minimum band gap lies. In the commensurate phase, the minimum band gap is at the boundaries ($k=0, \pi$) of the half Brillouin zone, while in the incommensurate phase, the minimum band gap lies within the half Brillouin zone \cite{Bunder1999prb}. In our model, the boundary of the commensurate-incommensurate transition is given by
\begin{equation}
    |h| = 1 - \gamma^{2}.
\end{equation}

\begin{figure}
    \centering
    \includegraphics[width=1\linewidth]{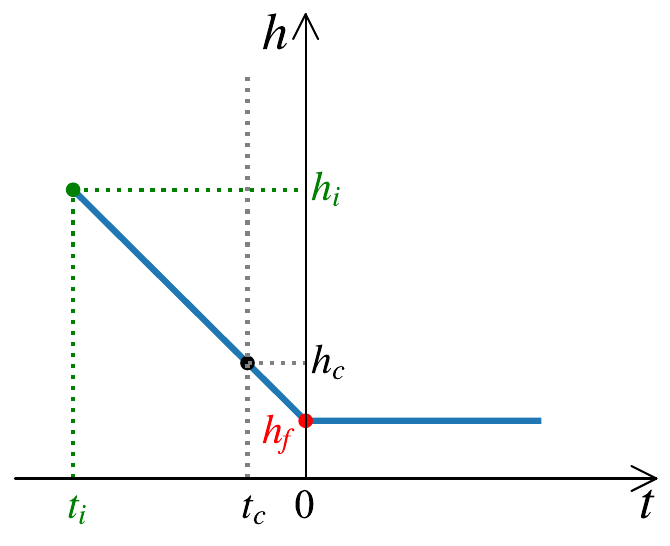}
    \caption{Illustration of the time-dependent transverse field $h(t)$ during and following a linear quench. $h(t)$ passes through the critical point $h_{c}=1$ at $t=t_{c}=(h_{f}-1)\tau_{Q}$.  }
    \label{fig:linear-quench}
\end{figure}

In the following, we consider the system is initially prepared in the ground state of the Hamiltonian $H(h_{i})$, i.e., $|\psi(t_{i})\rangle=|\mathrm{GS}_{i}\rangle$. Then, the transverse field is linearly changed to a final value $h_{f}$ at time $t=0$. Following the quench ($t>0$), the system evolves according to the Hamiltonian $H(h_{f})$. In other words, we consider the time-dependent system $H(t)=H[h(t)]$ with the continuous changing transverse field $h(t)$ taking (see Fig.~\ref{fig:linear-quench})
\begin{equation}
    h(t) = \left\{\begin{array}{cc}
                    -\frac{t}{\tau_{Q}}+h_{f}, & t<0, \\
                    h_{f}, & t>0.
                  \end{array}\right.
\end{equation}
During the quench, the system passes through the critical point $h_{c}$ at time $t_{c}=(h_{f}-1)\tau_{Q}$. It should be noted that in our model, we do not limit the annealing time to $\tau_{Q}$ as in Ref.~\onlinecite{Tatjana2016SciPost}, but instead, the duration of the ramp depends on the final field $h_{f}$ with a fixed initial field $h_{i}$.

In the ramp process ($t<0$), the dynamics of the system lie between two extreme limits: the sudden quench and the adiabatically slow change of parameters, so that one has the instantaneous Bogoliubov coefficients at each time. We assume the ansatz that the time-dependent state is also in the BCS-like form by
\begin{equation}\label{eq:state-during}
    |\psi(t)\rangle = \prod_{k>0}[u_{k}(t)+v_{k}(t)c_{k}^{\dag}c_{-k}^{\dag}]|0\rangle.
\end{equation}
The time-dependent coefficients $u_{k}(t)$ and $v_{k}(t)$ can be solved from the time-dependent Bogoliubov-de Gennes (BdG) equation
\begin{equation}\label{eq:BdG-Ising}
    i\frac{d}{dt}\left[
                   \begin{array}{c}
                     \tilde{u}_{k}(t) \\
                     \tilde{v}_{k}(t) 
                   \end{array}
                 \right]
    = \left[
        \begin{array}{cc}
          -h(t)-\cos{k} & -\gamma\sin{k} \\
          -\gamma\sin{k} & h(t)+\cos{k} 
        \end{array}
      \right]
      \left[
        \begin{array}{c}
          \tilde{u}_{k}(t) \\
          \tilde{v}_{k}(t) 
        \end{array}
      \right],
\end{equation}
where $\tilde{u}_{k}(t) = e^{i\pi/4}u_{k}(t)$ and $\tilde{v}_{k}(t) = e^{-i\pi/4}v_{k}(t)$ are obtained by applying the unitary transformation $U = \exp{(i\sigma^{z}\pi/4)}$. The equations can be solved to make them fully equivalent to the finite Landau-Zener problem \cite{Vitanov1996pra, Vitanov1999pra, zamani2310.15101}. 

After the quench, the system evolves according to the Hamiltonian $H(h_f)$ from the new initial state $|\psi(0)\rangle$, which is also the time-dependent state at the end of the ramp, i.e.,
\begin{equation}
    |\psi(0)\rangle = \prod_{k>0}[u_{k}(0)+v_{k}(0)c_{k}^{\dag}c_{-k}^{\dag}]|0\rangle.
\end{equation}
Considering the relation between the Jordan-Wigner fermions and the Bogoliubov fermions of the final field $h_{f}$
\begin{equation}
    c_{k} = \cos{\theta_{k}^{f}}\eta_{kf} - i\sin{\theta_{k}^{f}}\eta_{-kf}^{\dag},
\end{equation}
the time-dependent state following the quench ($t>0$) can be obtained by
\begin{equation}\label{eq:state.following}
    \begin{split}
        |\psi(t)\rangle & = e^{-iH(h_{f})t}|\psi(0)\rangle = \prod_{k>0} e^{-iH_{k}^{f}t}|\psi_{k}(0)\rangle \\
                        & = \prod_{k>0} \{[u_{k}(0)\cos{\theta_{k}^{f}}-iv_{k}(0)\sin{\theta_{k}^{f}}]e^{-i\varepsilon_{k}^{f}t}|0\rangle\rangle \\
                        & \quad\quad + [-iu_{k}(0)\sin{\theta_{k}^{f}}+v_{k}(0)\cos{\theta_{k}^{f}}]e^{i\varepsilon_{k}^{f}t}|1\rangle\rangle \},
    \end{split}
\end{equation}
where $\varepsilon_{k}^{f}$ is the quasiparticle excitation spectrum of $H(h_f)$, and $|0\rangle\rangle (|1\rangle\rangle=\eta_{kf}^{\dag}\eta_{-kf}^{\dag}|0\rangle\rangle)$ is the quasiparticle vacuum state of $H(h_f)$. Apparently, if the system follows the adiabatic evolution, the time-dependent state is exactly the instantaneous eigenstate, i.e. $|\psi(t) = \prod_{k}e^{-i\varepsilon_{k}^{f}t}|0\rangle\rangle$.

\section{Loschmidt echo and dynamical vortex}
\label{sec:Le}

The concept of the Loschmidt amplitude, also called return amplitude, measures the overlap between the time-evolved state $|\psi(t)\rangle = U(t)|\psi(t_{i})\rangle$ and the initial state $|\psi(t_{i})\rangle$, i.e.
\begin{equation}
    \mathcal{G}(t) = \langle\psi(t_{i})|\psi(t)\rangle = \langle\psi(t_{i}|U(t)|\psi(t_{i})\rangle.
\end{equation}
Due to the formal analogy between the Loschmidt amplitude and the canonical partition function \cite{Heyl2013prl}, one can define the dynamical free energy density by the rate function $\lambda(t) = -\lim_{N\rightarrow\infty}\frac{1}{N}\ln{|\mathcal{G}(t)|^{2}}$. The DQPT is characterized by the nonanalytic singularities of $\lambda(t)$ at specific critical times $t_{c}^{*}$. 

In the literature, we note that the dynamics of the Loschmidt amplitude induced by the slow quench have been considered previously \cite{Sharma2016prb, Divakaran2016pre, Tatjana2016SciPost, zamani2310}. Specifically, the Refs.~\cite{Sharma2016prb, Divakaran2016pre, zamani2310} solely considered the evolution following the ramp, i.e., the amplitude $\langle\psi(0)|\psi(t>0)\rangle$. The slow quench appears to be a way to prepare the ``initial state'' $|\psi(0)\rangle$. In these works, the ramp crossing the critical point is often a sufficient condition for a DQPT \cite{Heyl2018rpp}. In contrast, the Ref.~\onlinecite{Tatjana2016SciPost} consider the full-time evolution both during and after the ramp, i.e., $\langle\psi(t_{i})|\psi(t>t_{i})\rangle$. However, we will show that since they limited the duration of the ramp to the quench time $\tau_{Q}$, they only observed the DQPT for the slow quench (i.e. large $\tau_{Q}$).

\begin{figure}
    \centering
    \includegraphics[width=1\linewidth]{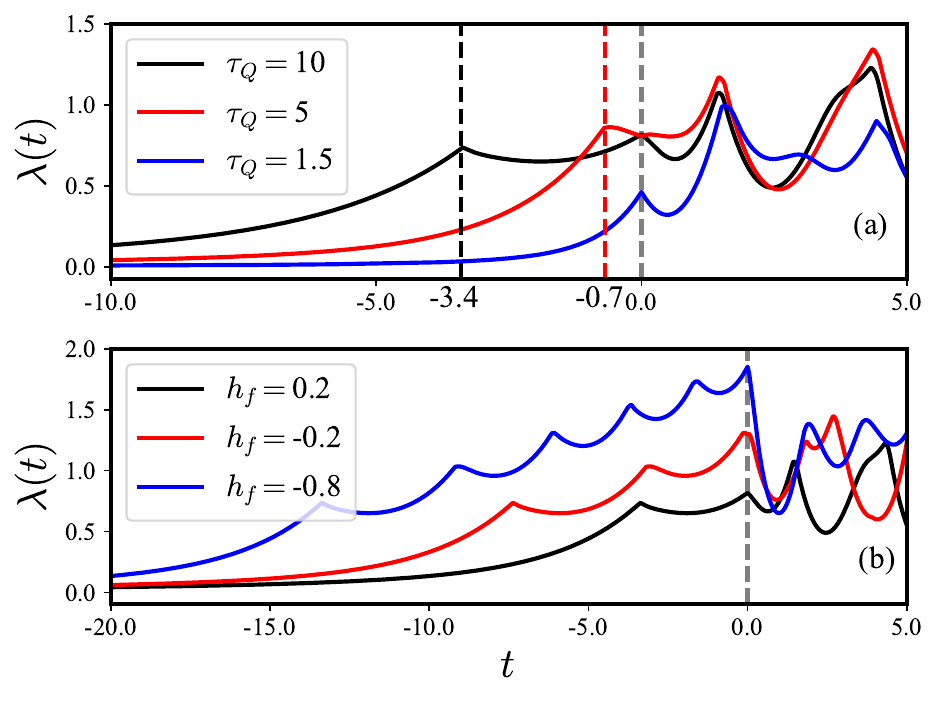}
    \caption{(a) Rate functions $\lambda(t)$ for ramps with varied quench times $\tau_{Q}=10$, $\tau_{Q}=5$, and $\tau_{Q}=1.5$. The linear quench is from $h_{i}=10$ to $h_{f}=0.2$ with the same anisotropic parameter $\gamma=0.8$. (b) Rate functions for the ramp with $\tau_{Q}=10$ from $h_{i}=10$ to varied final external fields $h_{f}=0.2$, $h_{f}=-0.2$, and $h_{f}=-0.8$. }
    \label{fig:rate_function}
\end{figure}

According to Eqs.~(\ref{eq:state-during}) and (\ref{eq:state.following}), we can immediately obtain the Loschmidt amplitude $\mathcal{G}(t) = \prod_{k>0}\mathcal{G}_{k}(t)$ with
\begin{equation}\label{eq:loschmidt_amplitude_during}
    \mathcal{G}_{k}(t) = u_{k}^{*}(t_{i})u_{k}(t)+v_{k}^{*}(t_{i})v_{k}(t),
\end{equation}
for $t<0$, and for $t>0$
\begin{equation}
    \mathcal{G}_{k}(t) = X_{k}e^{-i\varepsilon_{k}^{f}t} + Y_{k}e^{i\varepsilon_{k}^{f}t},
\end{equation}
where
\begin{equation*}
    \begin{split}
        X_{k} & = u_{k}^{*}(t_{i})u_{k}(0)\cos^{2}{\theta_{k}^{f}} + v_{k}^{*}(t_{i})v_{k}(0)\sin^{2}{\theta_{k}^{f}} \\
              & -i[u_{k}^{*}(t_{i})v_{k}(0)-v_{k}^{*}(t_{i})u_{k}(0)]\sin{\theta_{k}^{f}}\cos{\theta_{k}^{f}},
    \end{split}
\end{equation*}
and 
\begin{equation*}
    \begin{split}
        Y_{k} & = u_{k}^{*}(t_{i})u_{k}(0)\sin^{2}{\theta_{k}^{f}} + v_{k}^{*}(t_{i})v_{k}(0)\cos^{2}{\theta_{k}^{f}} \\
              & +i[u_{k}^{*}(t_{i})v_{k}(0)-v_{k}^{*}(t_{i})u_{k}(0)]\sin{\theta_{k}^{f}}\cos{\theta_{k}^{f}}.
    \end{split}
\end{equation*}
It is easy to observe the difference in the dynamic behavior of the Loschmidt amplitude during and following the quench. For the case after the ramp, the time $t$ appears in the exponent, i.e., $e^{\pm i\varepsilon_{k}^{f}t}$. This indicates that when the DQPT occurs, the critical times emerge periodically 
\begin{equation}
    t_{c}^{*}(n) = \frac{\varphi_{k^{*}}}{2\varepsilon_{k^{*}}^{f}} + t_{c}^{*}(0)(n+\frac{1}{2})
\end{equation}
where $t_{c}^{*}(0)=\frac{\pi}{\varepsilon_{k^{*}}^{f}}$, $k^{*}$ is the corresponding critical wave vector, and $\alpha_{k^{*}}$ is obtained by $\frac{Y_{k}}{X_{k}}=|\frac{Y_{k}}{X_{k}}|e^{i\varphi_{k^{*}}}$. However, according to Eq.~(\ref{eq:loschmidt_amplitude_during}), the critical times may not be spaced uniformly if there is more than one critical time.

\begin{figure}
    \centering
    \includegraphics[width=0.9\linewidth]{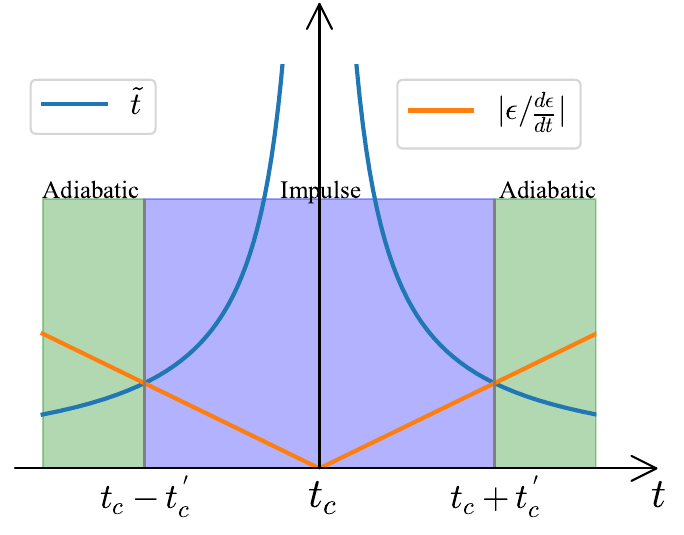}
    \caption{ The reaction time $\tilde{t}=|\epsilon|^{-\nu z}$ and the transition rate $|\epsilon/\frac{d\epsilon}{dt}|$ as a function of the time $t$, where $\epsilon=\frac{h-h_{c}}{h_{c}}$ is the dimensionless distance from the critical point $h_{c}$, diverges at the critical time $t_{c}$, and equals the transition rate at $t_{c} \pm t'_{c}$ with the frozen-out time $t'_{c}=\tau_{Q}^{\frac{\nu z}{1+\nu z}}$. In our work, we consider the case of the ramp across a critical point $h_{c}=1$, corresponding to $z=\nu=1$. } 
    \label{fig:AIA-approximation}
\end{figure}

In Fig.~\ref{fig:rate_function}~(a), we show three typical examples of the rate function corresponding to different $\tau_{Q}$. First of all, we observe the cusp-like peaks for all cases following the ramp ($t>0$). This means that the ramp crossing the critical points is still a sufficient condition to obtain a DQPT, although the initial state is changed to the beginning parameter of the linear quench. However, for cases during the quench ($t<0$), we see that the rate functions for slow quenches with $\tau_{Q}=10$ and $5$ exhibit a cusp-like peak, but for the rapid quench with $\tau_{Q}=1.5$ not. These results are consistent with those in Ref.~\onlinecite{Tatjana2016SciPost}, i.e., the DQPT may be absent when the linear quench is getting faster. Notably, these rate functions consistently show a cusp-like peak at the endpoints of the linear quenches ($t=0$). We will explain later that this non-analytic point is due to the different dynamics processes before and after, rather than corresponding to the zero point of the Loschmidt echo.

To unravel this phenomenon, we turn to the adiabatic-impulse-adiabatic (AIA) approximation \cite{Dziarmaga2010advp, Kou2023prb}. In this approximation scheme, depicted in Fig.~\ref{fig:AIA-approximation}, the system is initially prepared in the ground state at $t_{i} \rightarrow -\infty$, ensuring a sizable gap between the ground state and the first relevant excited state. The rapid reaction time $\tilde{t}$ allows the system to follow its adiabatic ground state until $t_{c}-t'_{c}$, where $t'_{c}=\tau_{Q}^{\frac{\nu z}{1+\nu z}}$ is the frozen-out time. Consequently, the occurrence of a DQPT is precluded during the left adiabatic regime (see Fig.~\ref{fig:AIA-approximation}). In principle, the appearance of a DQPT requires a time of the ramp at least $t \gtrsim t_{c}-t'_{c}$, but in reality, we have found through calculations that the time for the occurrence of DQPT requires $t>t_{c}$. In our XY chain, the frozen-out time is given by $t'_{c} = \tau_{Q}^{1/2}$ as $\nu = z = 1$. Specifically, in the case depicted in Fig.~\ref{fig:rate_function}~(a), a DQPT can only emergent when $t \gtrsim t_{c}-t'_{c} \approx -11.16$ for $\tau_{Q}=10$, $t \gtrsim -6.24$ for $\tau_{Q}=5$, and $t \gtrsim -2.42$ for $\tau_{Q}=1.5$. Evidently, in the scenario with $\tau_{Q}=1.5$, the time available for the system to undergo a DQPT is insufficient.

Given that the timing of DQPT occurrences hinges on the adiabatic regime duration, it is conceivable that an increase in the remaining time $(t_{c}-t'_{c}<t<0)$ may lead to the manifestation of more DQPTs. To substantiate this inference, we analyze the rate functions for linear quenching processes with varying $h_{f}$ but the same quench time $\tau_{Q}$, as shown in Fig.~\ref{fig:rate_function}~(b). In this instance, $t_{c}-t'_{c}$ assumes values of $-11.16$ for $h_{f}=0.2$, $-15.16$ for $h_{f}=-0.2$, and $-21.16$ for $h_{f}=-0.8$. Consequently, we observe five cusp-like peaks for the linear quench towards $h_{f}=-0.8$, and three peaks for $h_{f}=-0.2$. Moreover, unlike the case following the ramp, the critical times of the DQPTs during the quench are not spaced periodically.

\begin{figure}
    \centering
    \includegraphics[width=1\linewidth]{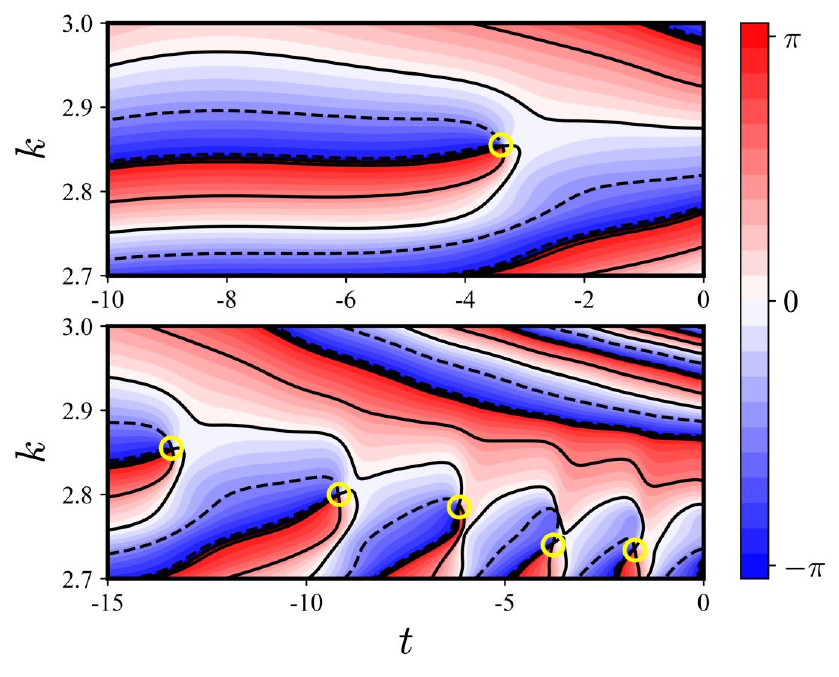}
    \caption{The contour plot of the geometric phase $\phi^{G}(k,t)$ in the $(t, k)$ plane. The linear quenches with $\tau_{Q}=10$ are from $h_{i}=10$ to $h_{f}=0.2$ for the upper panel, and from $h_{i}=10$ to $h_{f}=-0.8$ for the lower panel. The dynamical vortices are marked in yellow. }
    \label{fig:geometric.phase}
\end{figure}

The occurrence of a DQPT is actually a signature of the appearance, movement, and annihilation of dynamical vortices in time-momentum space \cite{Flaschner2018natphys14, Yu2017pra96, Qiu2018pra98, Lahiri2019prb99}. In order to establish connections between DQPTs and the theory of dynamical vortices, a dynamical topological order parameter (DTOP) has been proposed and extensively studied \cite{Budich2016prb93, Bhattacharya2017prb96, Bhattacharjee2018prb97, zhou2018pra98, Lang2018prb98, Yang2018prb97, Jafari2021pra103, cao2023prb108, Cao2024jpcm36}. The Loschmidt amplitude is written as $\mathcal{G}_{k}(t) = r_{k}(t)e^{i\phi(k,t)}$ in the polar coordinate, where the phase $\phi(k,t) = \phi^{\mathrm{G}}(k,t) + \phi^{\mathrm{dyn}}(k,t)$ contains two components---the geometric phase $\phi^{\mathrm{G}}(k,t)$ and the dynamical phase $\phi^{\mathrm{dyn}}(k,t)$. The dynamical phase $\phi^{\mathrm{dyn}}(k,t)$ depends on the instantaneous eigenenergy varying with time, which is defined as
%\begin{equation}
%    \phi^{\mathrm{dyn}}(k,t) = -\int_{0}^{t}\varepsilon_{k}(s)ds,
%\end{equation}
%where $\varepsilon_{k}(s) = \sqrt{[h(s)+\cos{k}]^{2}+\gamma^{2}\sin^{2}{k}}$.
\begin{equation}
    \phi^{\mathrm{dyn}}(k,t) = -\int_{0}^{t}ds\langle\psi_{k}(s)|H_{k}(s)|\psi_{k}(s)\rangle,
\end{equation}
where the time-dependent state is
\begin{equation}
    \begin{split}
        |\psi_{k}(s)\rangle & = u_{k}(s)|0\rangle + v_{k}(s)|1\rangle \\
                            & = [u_{k}(s)\cos{\theta_{k}(s)}-iv_{k}(s)\sin{\theta_{k}(s)}]|0\rangle\rangle \\
                            & + [-iu_{k}(s)\sin{\theta_{k}(s)}+v_{k}(s)\cos{\theta_{k}(s)}]|1\rangle\rangle \\
                            & = P_{k}(s)|0\rangle\rangle + Q_{k}(s)|1\rangle\rangle
    \end{split}
\end{equation}
the instantaneous Hamiltonian is 
\begin{equation}
    H_{k}(s) = \varepsilon_{k}(s)(\eta_{k}^{\dag}\eta_{k}-\frac{1}{2}) + \varepsilon_{-k}(s)(\eta_{-k}^{\dag}\eta_{-k}-\frac{1}{2}),
\end{equation}
and $\varepsilon_{k}(s) = \sqrt{[h(s)+\cos{k}]^{2}+\gamma^{2}\sin^{2}{k}}$. From $\eta_{k}^{\dag}\eta_{k}|0\rangle\rangle=0$ and $\eta_{-k}^{\dag}\eta_{-k}|0\rangle\rangle=0$, we have
\begin{equation}
    \phi^{\mathrm{dyn}}(k,t) = -\int_{0}^{t} \varepsilon_{k}(s)(|Q_{k}(s)|^{2}-|P_{k}(s)|^{2})ds.
\end{equation}
Consequently, we obtain the geometric phase $\phi^{G}(k,t)$ by
\begin{equation}
    \phi^{G}(k,t) = \mathrm{Arg}[\mathcal{G}_{k}(t)]-\phi^{\mathrm{dyn}}(k,t).
\end{equation}

\begin{figure}
    \centering
    \includegraphics[width=1\linewidth]{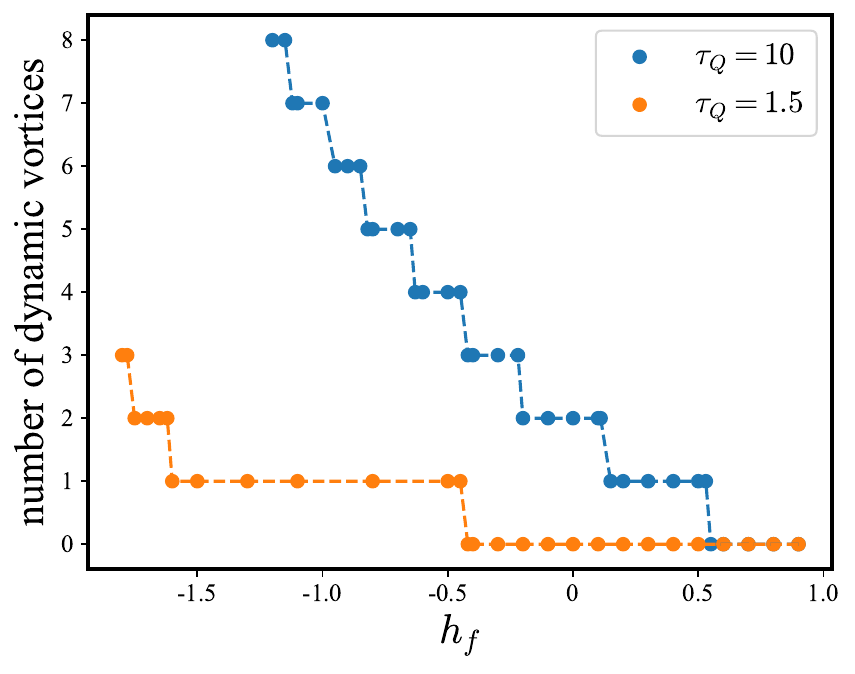}
    \caption{Number of the dynamical vortices as a function of the final external field $h_{f}$ of the linear quench for the quench times $\tau_{Q}=10$, and $\tau_{Q}=1.5$. The initial parameter of the linear quench is $(h_{i}=10, \gamma_{i}=0.8)$. }
    \label{fig:num_vortices}
\end{figure}

The contour plots of geometric phases $\phi^{G}(k,t)$ for linear quenches from $h_{i}=10$ to $h_{f}=0.2$, and from $h_{i}=10$ to $h_{f}=-0.8$ are shown in Fig.~\ref{fig:geometric.phase}. We observe that dynamical vortices are emergent at critical times. In the lower panel of Fig.~\ref{fig:geometric.phase}, the critical vectors corresponding to each critical time are not the same, which is the reason why the critical times during the ramp are not spaced uniformly \cite{Cao2024jpcm36}. Notably, we also notice that there is no dynamical vortex at time $t=0$. This confirms that the cusp-like peak (see Fig.~\ref{fig:rate_function}) of rate functions at $t=0$ does not originate from the nonanalytic behavior of the Loschmidt amplitude, but is caused by the distinct dynamics processes preceding and following the time $t=0$.

Furthermore, since the critical times are not spaced periodically during the quench, the number of dynamical vortices is actually dependent on the parameters of the final Hamiltonian. Fig.~\ref{fig:num_vortices} illustrates the variation in the number of dynamical vortices with the final external field $h_{f}$ for quench times $\tau_{Q}=10$ and $1.5$. The number of dynamical vortices during the linear quench ($t<0$) depends on the final external field $h_{f}$ indicating a different scenario of DQPT compared to dynamical vortices emerging periodically \cite{Budich2016prb93}. This result also supports the previous finding that, for a faster quench, achieving a DQPT requires quenching to a smaller magnetic field to ensure adequate time for the emergence of dynamical vortices. In other words, the quench crossing the critical point during the linear quench is not a sufficient condition for a present DQPT.

\section{Dynamics of the fermion two-point correlation function}
\label{sec:order-parameter}

\subsection{Time-dependent part of the fermion two-point correlation function}
\label{sec:relaxation}

In this section, we investigate the DPT from the long-term perspective, in which the different relaxation dynamics are used to define the dynamical phase. The DQPT previously describes the asymptotic long-time steady state of order parameters varying with the system parameters \cite{Prosen2008prl, Barmettler2009prl, Yuzbashyan2006prl, Sciolla2010prl}. However, recent works have suggested that the asymptotic scaling behavior of order parameters approaching their steady values can also characterize dynamical phases. This viewpoint was initially studied in periodically driven systems \cite{Sen2016prb, Sourav2018jpa, Sarkar2020prb, Aditya2022prb} and later extended to the quantum quench protocols \cite{Makki2022prb, Zou2023prb, Cao2024prb109, Ramos2023prb}. In our work, we focus on the fermionic two-point correlation function $C_{mn}(t) = \langle\psi(t)|c_{m}^{\dag}c_{n}|\psi(t)\rangle$ following the Ref.~\onlinecite{Sen2016prb, Makki2022prb, Cao2024prb109}. 

\begin{figure}
    \centering
    \includegraphics[width=1\linewidth]{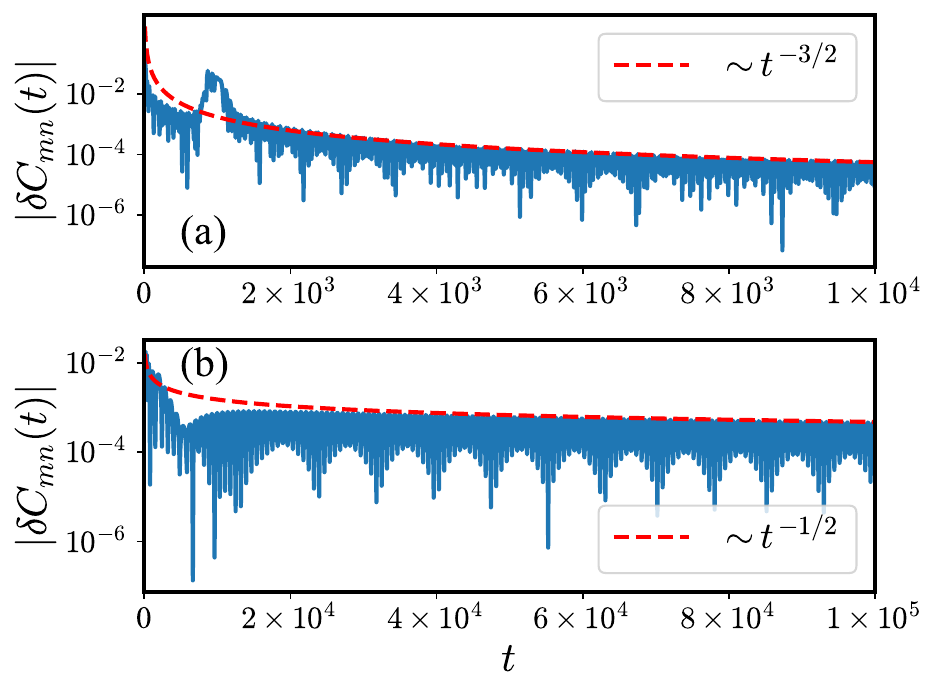}
    \caption{$|\delta C_{mn}(t)|$ as a function of $t$ after the ramp with $\tau_{Q}=10$. In the upper panel, the linear quench is from $(h_{i}=10, \gamma_{i}=1.0)$ to $(h_{f}=0.1, \gamma_{f}=1.0)$, corresponding to the case to the commensurate phase. In the lower panel, the linear quench is from  $(h_{i}=10, \gamma_{i}=0.5)$ to $(h_{f}=0.1, \gamma_{f}=0.5)$, corresponding to the case to the incommensurate phase.}
    \label{fig:delta_C_KZ}
\end{figure}

In the dynamical protocol we study, the system undergoes an initial finite-time linear quench followed by evolution under the final Hamiltonian $H(h_f)$. The fermion two-point correlation function $C_{mn}(t)$ approaches a steady value following the linear quench ($t>0$). According to Eq.~(\ref{eq:state.following}), we obtain the expression of $C_{mn}(t)$ by
\begin{equation}
    \begin{split}
        C_{mn}(t) & = C_{mn}(\infty) + \delta C_{mn}(t) \\
                  & = \frac{1}{N}\sum_{k>0} C_{mn}(k, \infty) + \delta C_{mn}(k, t)
    \end{split} 
\end{equation}
with
\begin{equation}
    \begin{split}
        C_{mn}(k, \infty) & =  [r_{k}^{2}+s_{k}^{2}+\frac{1}{2}(p_{k}^{2}+q_{k}^{2}-r_{k}^{2}-s_{k}^{2})\sin^{2}{2\theta_{k}^{f}} \\
        & -\frac{1}{2}(p_{k}s_{k}-q_{k}r_{k})\sin{4\theta_{k}^{f}}]\cos{k(n-m)}
    \end{split}
\end{equation}
and 
\begin{equation}
    \begin{split}
        \delta C_{mn}(k, t) & = [(q_{k}r_{k}-p_{k}s_{k})\sin{4\theta_{k}^{f}} \\
        & +(p_{k}^{2}+q_{k}^{2}-r_{k}^{2}-s_{k}^{2})\sin^{2}{2\theta_{k}^{f}}]\cos{(2\varepsilon_{k}^{f}t)} \\
        & - 2(p_{k}r_{k}+q_{k}s_{k})\sin{2\theta_{k}^{f}}\sin{(2\varepsilon_{k}^{f}t)},
    \end{split}
\end{equation}
where the functions $p_{k}$, $q_{k}$, $r_{k}$, and $s_{k}$ are given by $u_{k}(0) = p_{k}+iq_{k}$ and $v_{k}(0) = r_{k}+is_{k}$. Similar to the case in the sudden quench, the relaxation behavior of $\delta C_{mn}(t)$ can be obtained by the stationary point approximation, i.e.,
\begin{equation}
    \delta C_{mn}(t) = \int_{0}^{\pi}\frac{dk}{2\pi}\{\mathrm{Re}[f(k)e^{2i\varepsilon_{k}^{f}t}]+\mathrm{Im}[g(k)e^{2i\varepsilon_{k}^{f}t}]\}.
\end{equation}
Consequently, we find that the scaling behavior of $\delta C_{mn}(t)$ only depends on the stationary behavior of the quasiparticle excitation spectra $\varepsilon_{k}^{f}$, which is the way to determine the commensurate and incommensurate phases, irrespective of the details of the quench protocols. In Fig.~\ref{fig:delta_C_KZ}, we display the typical examples of the numerical results for the ramp with $\tau_{Q}=10$ to the commensurate and incommensurate phases, respectively. It is observed that the deviation $\delta C_{mn}(t)$ follows the scaling law behavior of $\sim t^{-3/2}$ for the ramp to the commensurate phase, and $\sim t^{-1/2}$ for the ramp to the incommensurate phase. 

To gain more physical insights, we try to analyze the relaxation behavior of $\delta C_{mn}(t)$ via the AIA approximation as shown in Fig.~\ref{fig:AIA-approximation}. The system is initially prepared in the ground state at $t_{i}\rightarrow\infty$, where the gap between the ground state and the first relevant excited state is large enough. The reaction time $\tilde{t}$ is also fast enough so that the state of the system can follow its adiabatic ground state until the time $t_{c}-t'_{c}$. Since the ramp with $\tau_{Q}\gg1$ has a narrow impulse region and the reaction time increases very fast in it (see Fig.~\ref{fig:AIA-approximation}), the system reacts too slowly to follow the evolving external field in the impulse region. Hence the state is effectively frozen out. The adiabatic evolution of the state restarts again at $t_{c}+t'_{c}$. In this approximation, the ramp in the impulse stage is equivalent to a sudden quench from $h_{c}(1+\epsilon_{c})$ to $h_{c}(1-\epsilon_{c})$. As a result, the state at time $t_{c}-t'_{c}$ and $t_{c}+t'_{c}$ are
\begin{equation}
    \begin{split}
        |\psi_{k}(t_{c}+t'_{c})\rangle & = |\psi_{k}(t_{c}-t'_{c})\rangle = |0\rangle\rangle_{+} \\
                                       & = \cos{\alpha_{k}}|0\rangle\rangle_{-}-i\sin{\alpha_{k}}|1\rangle\rangle_{-},
    \end{split}
\end{equation}
where $\alpha_{k} = \theta_{k}^{-}-\theta_{k}^{+}$ denotes the difference between the Bogoliubov angles $\theta_{k}^{+}$ and $\theta_{k}^{-}$ of the system in the external fields $h_{c}(1+\epsilon_{c})$ and $h_{c}(1-\epsilon_{c})$, and $|0\rangle\rangle_{+}$, $|0\rangle\rangle_{-}$ denote the corresponding quasiparticle vacuum states. Notably, the state $|\psi_{k}(t_{c}+t'_{c})\rangle$ at $t_{c}+t'_{c}$ becomes the initial state for the following adiabatic evolution. Therefore, the time-evolved state after the ramp in the AIA approximation is given by
\begin{equation}
    |\Psi(t)\rangle = \prod_{k>0}\cos{\alpha_{k}}e^{-i\varepsilon_{k}^{f}t}|0\rangle\rangle -i\sin{\alpha_{k}}e^{i\varepsilon_{k}^{f}t}|1\rangle\rangle. 
\end{equation}
Then we can immediately obtain the derivation of $\delta C_{mn}(t)$ following the Ref.~\onlinecite{Cao2024prb109}, i.e.,
\begin{equation}
    \delta C_{mn}(k, t) = \sin{2\theta_{k}^{f}}\sin{2\alpha_{k}}\cos{2\varepsilon_{k}^{f}t}\cos{[k(n-m)]}. 
\end{equation}
In the AIA approximation, the relaxation behavior of $\delta C_{mn}(t)$ is actually equivalent to the case of quench from the gapped phase \cite{Cao2024prb109}. 
 
\subsection{Steady part of the fermion two-point correlation function}
\label{sec:steady}

\begin{figure}
    \centering
    \includegraphics[width=1\linewidth]{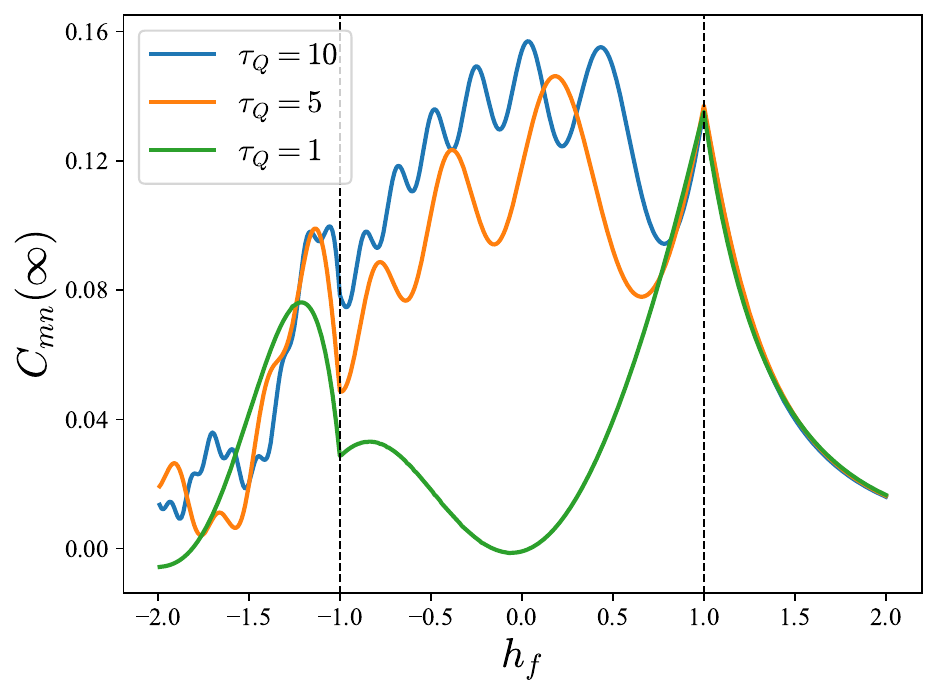}
    \caption{Steady value $C_{mn}(\infty)$ of the fermion two-point correlation function as a function of the final external filed $h_{f}$ for the linear quenches. The linear quench is from $h_{i}=10$ with the fixed anisotropic parameter $\gamma=0.8$. }
    \label{fig:steady_value_hf}
\end{figure}

In addition to discussing the asymptotic behavior of the time-dependent part $\delta C_{mn}(t)$, we also focus on the steady part of the fermion two-point correlation function. In Ref.~\onlinecite{cao2024relaxation}, we have found that although the steady value $C_{mn}(\infty)$ is different from the ground state expectation value, it can also be used to characterize the ground state quantum phase transition (QPT). Here, we confirm this conclusion in the XY chain following the linear quench. Fig.~\ref{fig:steady_value_hf} displays several typical examples of the steady value $C_{mn}(\infty)$ varying with the final external field $h_{f}$. We observe that $C_{mn}(\infty)$ exhibits nonanalytic cusp-like singularities at the critical points $h_{c}=\pm1$. The results further support the potential use of the steady value $C_{mn}(\infty)$ as a signature of QPTs. 

In our model, this conclusion can be understood through extreme cases. When the linear quench is very slow, the behavior of the system will be very close to the values of the ground state, even if the quench crosses the critical point. In particular, we note that the Ref.~\onlinecite{Tatjana2016SciPost} has found the steady values approaching to the ground-state expectation value with a power-law relationship $\sim \tau_{Q}^{-1/2}$ in the adiabatic limit.

\section{Conclusion}
\label{sec:conclusion}

In this study, we investigate the nonequilibrium dynamics induced by the finite-time linear quench in the XY chain. Initially, we examine the DQPT, characterized by the nonanalytic behavior of the Loschmidt amplitude. We find that following the ramp, the quench crossing the critical point $h_{c}$ is a sufficient condition for the occurrence of the DQPT, and the critical times are spaced periodically. It is noteworthy that while our initial state differs from previous studies \cite{Sharma2016prb, Divakaran2016pre, zamani2310}, the ultimate conclusion remains consistent. However, crossing the critical point during the ramp is no longer a sufficient condition, and the critical times are not spaced periodically. Through the AIA approximation analysis, we deduce that critical times must follow the time for the quench crossing the critical point, implying that the occurrence of DQPT needs adequate time after crossing the critical point. Therefore, the manifestation of DQPT necessitates adequate distancing of the quenched parameters from the critical point. This is why the Ref.~\onlinecite{Tatjana2016SciPost} finds the DQPT is absent as the linear quench gets faster. Furthermore, we corroborate our conclusions by computing the dynamical vortices. 

Conversely, we explore the relaxation dynamics of the fermion two-point correlation function, whose scaling behavior transition is proposed as a distinct type of DPT. By deriving the expression of $\delta C_{mn}(t)$, we establish that the asymptotic behavior of $\delta C_{mn}(t)$ is dependent on the stationary behavior of the quasiparticle excitation spectra $\varepsilon_{k}^{f}$ of the final Hamiltonian. Consequently, we observe that the relaxation dynamics of $\delta C_{mn}(t)$ are contingent upon whether the final Hamiltonian resides in the commensurate phase or the incommensurate phase, with $\delta C_{mn}(t)$ exhibiting power-law behavior of $\sim t^{-3/2}$ and $\sim t^{-1/2}$ for ramps to the commensurate and incommensurate phases, respectively. Moreover, in the AIA approximation, the relaxation behavior following the linear quench is equivalent to that of a sudden quench. Additionally, we delve into the steady part of the fermion two-point correlation function, where the steady value $C_{mn}(t)$ showcases nonanalytic singularities at the critical point. 

In summary, our study delves into two distinct types of dynamical phase transitions, examining them from short-term and long-term perspectives. By shedding light on these phenomena, our research contributes to a deeper understanding of quantum phase transitions in nonequilibrium systems.

\begin{acknowledgments}
    The work is supported by the National Science Foundation of China (Grant Nos.~11975126 and 12247106).
\end{acknowledgments}

\appendix
\begin{widetext}

\section{Solving time-dependent BdG equation}
\label{sec:BdG}

In our model, the time-dependent BdG equations can be solved by mapping to the finite Landau-Zener problem \cite{Vitanov1996pra, Vitanov1999pra}. By defining 
\begin{eqnarray}
    \bar{t} &=& [t-(h_{f}+\cos{k})\tau_{Q}]\gamma\sin{k}, \\
    \tau &=& \tau_{Q}\gamma^{2}\sin^{2}{k},
\end{eqnarray}
and 
\begin{equation}
    \left[\begin{array}{c}
            \tilde{u}_{k}(\bar{t}) \\
            \tilde{v}_{k}(\bar{t}) 
          \end{array}
    \right] 
    =
    \left[\begin{array}{c}
            \tilde{u}_{k}(t) \\
            \tilde{v}_{k}(t) 
          \end{array}
    \right], 
\end{equation}
we obtain the exact Landau-Zener equations
\begin{equation}\label{eq:LZ-equation}
    i\frac{d}{d\bar{t}}\left[\begin{array}{c}
                              \tilde{v}_{k}(\bar{t}) \\
                              \tilde{u}_{k}(\bar{t}) 
                            \end{array}
                        \right]
    = \left[\begin{array}{cc}
              -\alpha\bar{t} & 1 \\
              1 & \alpha\bar{t} \\
            \end{array}
      \right]\left[\begin{array}{c}
                    \tilde{v}_{k}(\bar{t}) \\
                    \tilde{u}_{k}(\bar{t}) 
                  \end{array}
            \right]    
\end{equation}
with $\alpha = 1/\tau$. Note that the order of $\tilde{u}_{k}(\bar{t})$ and $\tilde{v}_{k}(\bar{t})$ is changed.
The exact solution of Eq.~(\ref{eq:LZ-equation}) is given by (see also in Ref.~\onlinecite{zamani2310.15101})
\begin{equation}
    \left[\begin{array}{c}
            \tilde{v}_{k}(\bar{t}) \\
            \tilde{u}_{k}(\bar{t}) 
          \end{array}
    \right] = \left[\begin{array}{cc}
                      U_{11}(z) & U_{12}(z) \\
                      U_{21}(z) & U_{22}(z) \\
                    \end{array}
              \right]
              \left[\begin{array}{c}
                      \tilde{v}_{k}(\bar{t}_{i}) \\
                      \tilde{u}_{k}(\bar{t}_{i}) 
                    \end{array}
              \right]
\end{equation}
with
\begin{equation}
    U_{11}(z) = \frac{\Gamma(1-p)}{\sqrt{2\pi}}[D_{p-1}(-z_{i})D_{p}(z) + D_{p-1}(z_{i})D_{p}(-z)],
\end{equation}
\begin{equation}
    U_{12}(z) = \frac{\Gamma(1-p)}{\omega\sqrt{\pi}}e^{i\pi/4}[D_{p}(z_{i})D_{p}(-z) - D_{p}(-z_{i})D_{p}(z)],
\end{equation}
\begin{equation}
    \begin{split}
        U_{21}(z) & = \frac{\omega\Gamma(1-p)}{2\sqrt{\pi}}e^{-i\pi/4}[D_{p-1}(z_{i})D_{p-1}(-z) \\
                  & \quad\quad\quad\quad\quad\quad\quad\quad - D_{p-1}(-z_{i})D_{p-1}(z)],
    \end{split}
\end{equation}
and 
\begin{equation}
    U_{22}(z) = \frac{\Gamma(1-p)}{\sqrt{2\pi}}[D_{p}(-z_{i})D_{p-1}(z) + D_{p}(z_{i})D_{p-1}(-z)],
\end{equation}
where $z = e^{-i\pi/4}\sqrt{2\alpha}\bar{t}$, $z_{i} = e^{-i\pi/4}\sqrt{2\alpha}\bar{t}_{i}$, $p=i/2\alpha$, $\omega=1/\sqrt{\alpha}$, and $D_{p}(x)$ is the parabolic cylinder function \cite{Olver2010NISTHO}. $D_{p}(x)$ has following asymptotic expansions for $|x| \gg 1$,
\begin{equation}
    D_{p}(x) = e^{-x^{2}/4}x^{p}[1+o(x^{-2})], \quad \forall |\mathrm{arg}(x)| < 3\pi/4,
\end{equation}
and for $|x| \ll 1$,
\begin{equation}
    D_{p}(x) = \frac{2^{\frac{p}{2}}\sqrt{\pi}}{\Gamma(\frac{1}{2}-\frac{p}{2})} - \frac{2^{\frac{1}{2}+\frac{p}{2}}\sqrt{\pi}x}{\Gamma(-\frac{p}{2})} + o(x^{2}).
\end{equation}
To find other values of $\mathrm{arg}(z)$ of the parabolic cylinder functions, one can use the connection formula such as
\begin{equation}
    D_{p}(-x) = e^{i\pi x}D_{p}(x) + \frac{\sqrt{2\pi}}{\Gamma(-p)}e^{i(p+1)\pi/2}D_{-p-1}(ix),
\end{equation}
due to $\mathrm{arg}(ix)=\mathrm{arg}(x)+\pi/2$ and $\mathrm{arg}(-x)=\mathrm{arg}(x)+\pi$.

%The time-dependent BdG equation for our model is expressed as 
%\begin{equation}
%    i\frac{d}{dt}\left[\begin{array}{c}
%                         u_{k}(t) \\
%                         v_{k}(t) 
%                       \end{array}\right]
%                = \left[
%                    \begin{array}{cc}
%                      -h(t)-\cos{k} & i\gamma\sin{k} \\
%                      -i\gamma\sin{k} & h(t)+\cos{k} 
%                    \end{array}
%                  \right]
%                  \left[\begin{array}{c}
%                         u_{k}(t) \\
%                         v_{k}(t) 
%                       \end{array}
%                  \right].
%\end{equation}
%By defining 
%\begin{equation}
%    u_{k}(t) = p_{k}(t)+iq_{k}(t), \quad v_{k}(t) = r_{k}(t)+is_{k}(t),
%\end{equation}
%we obtain
%\begin{eqnarray}
%    \frac{d}{dt}p_{k}(t) &=& [-h(t)-\cos{k}]q_{k}(t) - r_{k}(t)\sin{k}, \\
%    \frac{d}{dt}q_{k}(t) &=& [h(t)+\cos{k}]p_{k}(t) - s_{k}(t)\sin{k}, \\
%    \frac{d}{dt}r_{k}(t) &=& p_{k}(t)\sin{k} + [h(t)+\cos{k}]s_{k}(t), \\
%    \frac{d}{dt}s_{k}(t) &=& q_{k}(t)\sin{k} - [h(t)+\cos{k}]r_{k}(t)
%\end{eqnarray}
%with the initial condition $p_{k}(t_{i}) = \cos{\theta_{k}^{i}}, q_{k}(t_{i}) = 0, r_{k}(t_{i}) = 0, s_{k}(t_{i}) = \sin{\theta_{k}^{i}}$, where $\theta_{k}^{i}$ is the Bogoliubov angle of the initial Hamiltonian.

\section{Fermion correlation function after the ramp}

To derive the expression of $C_{mn}(t)$, we transform the fermionic operators in the momentum space, given by
\begin{equation}
    C_{mn}(t) = \langle\Psi(t)|\frac{1}{N}\sum_{k}c_{k}^{\dag}c_{k}e^{ik(n-m)}|\Psi(t)\rangle = \frac{1}{N}\sum_{k>0}[\langle\Psi_{k}(t)|c_{k}^{\dag}c_{k}|\Psi_{k}(t)\rangle e^{ik(n-m)} 
                +\langle\Psi_{k}(t)|c_{-k}^{\dag}c_{-k}|\Psi_{k}(t)\rangle e^{-ik(n-m)}],
\end{equation}
where the fermionic operators can be transformed by the quasiparticle operators 
\begin{equation}\label{eq:ck}
    \begin{split}
        \langle\Psi_{k}(t)|c_{k}^{\dag}c_{k}|\Psi_{k}(t)\rangle & = \langle\Psi_{k}(t)| (\cos{\theta_{k}^{f}}\eta_{k}^{\dag}+i\sin{\theta_{k}^{f}}\eta_{-k}) (\cos{\theta_{k}^{f}}\eta_{k}-i\sin{\theta_{k}^{f}}\eta_{-k}^{\dag}) |\Psi_{k}(t)\rangle \\
            & = \langle\Psi_{k}(t)| \cos^{2}{\theta_{k}^{f}}\eta_{k}^{\dag}\eta_{k} - i\sin{\theta_{k}^{f}}\cos{\theta_{k}^{f}}\eta_{k}^{\dag}\eta_{-k}^{\dag} + i\sin{\theta_{k}^{f}}\cos{\theta_{k}^{f}}\eta_{-k}\eta_{k} + \sin^{2}{\theta_{k}^{f}}\eta_{-k}\eta_{-k}^{\dag} |\Psi_{k}(t)\rangle,
    \end{split}
\end{equation}
and 
\begin{equation}\label{eq:c-k}
    \begin{split}
        \langle\Psi_{k}(t)|c_{-k}^{\dag}c_{-k}|\Psi_{k}(t)\rangle & = \langle\Psi_{k}(t)| (\cos{\theta_{k}^{f}}\eta_{-k}^{\dag}-i\sin{\theta_{k}^{f}}\eta_{k}) (\cos{\theta_{k}^{f}}\eta_{-k}+i\sin{\theta_{k}^{f}}\eta_{k}^{\dag}) |\Psi_{k}(t)\rangle \\
            & = \langle\Psi_{k}(t)| \cos^{2}{\theta_{k}^{f}}\eta_{-k}^{\dag}\eta_{-k} + i\sin{\theta_{k}^{f}}\cos{\theta_{k}^{f}}\eta_{-k}^{\dag}\eta_{k}^{\dag} - i\sin{\theta_{k}^{f}}\cos{\theta_{k}^{f}}\eta_{k}\eta_{-k} + \sin^{2}{\theta_{k}^{f}}\eta_{k}\eta_{k}^{\dag} |\Psi_{k}(t)\rangle.
    \end{split}
\end{equation}
Hence, we need to calculate every component in Eqs.~(\ref{eq:ck}) and (\ref{eq:c-k}). We obtain
\begin{equation}
    \begin{split}
        \langle\Psi_{k}(t)|\eta_{k}^{\dag}\eta_{k}|\Psi(t)\rangle & = 1 - \langle\Psi_{k}(t)|\eta_{k}\eta_{k}^{\dag}|\Psi_{k}(t)\rangle \\
        & = |u_{k}(0)|^{2}\sin^{2}{\theta_{k}^{f}} + |v_{k}(0)|^{2}\cos^{2}{\theta_{k}^{f}} + i[u_{k}^{*}(0)v_{k}(0)-u_{k}(0)v_{k}^{*}(0)]\sin{\theta_{k}^{f}}\cos{\theta_{k}^{f}},
    \end{split}
\end{equation}
\begin{equation}
    \begin{split}
        \langle\Psi_{k}(t)|\eta_{k}^{\dag}\eta_{-k}^{\dag}|\Psi_{k}(t)\rangle & = -\langle\Psi_{k}(t)|\eta_{-k}^{\dag}\eta_{k}^{\dag}|\Psi_{k}(t)\rangle \\
        & = \{u_{k}^{*}(0)v_{k}(0)\sin^{2}{\theta_{k}^{f}} + v_{k}^{*}(0)u_{k}(0)\cos^{2}{\theta_{k}^{f}} + i[|u_{k}(0)|^{2}-|v_{k}(0)|^{2}]\sin{\theta_{k}^{f}}\cos{\theta_{k}^{f}}\}e^{-2i\varepsilon_{k}^{f}t},
    \end{split}
\end{equation}
\begin{equation}
    \begin{split}
        \langle\Psi_{k}(t)|\eta_{-k}\eta_{k}|\Psi_{k}(t)\rangle & = -\langle\Psi_{k}(t)|\eta_{k}\eta_{-k}|\Psi_{k}(t)\rangle \\
        & = \{u_{k}^{*}(0)v_{k}(0)\cos^{2}{\theta_{k}^{f}} + v_{k}^{*}(0)u_{k}(0)\sin^{2}{\theta_{k}^{f}} + i[|v_{k}(0)|^{2}-|u_{k}(0)|^{2}]\sin{\theta_{k}^{f}}\cos{\theta_{k}^{f}}\}e^{2i\varepsilon_{k}^{f}t},
    \end{split}
\end{equation}
\begin{equation}
    \begin{split}
        \langle\Psi_{k}(t)|\eta_{-k}\eta_{-k}^{\dag}|\Psi_{k}(t)\rangle & = 1 - \langle\Psi_{k}(t)|\eta_{-k}^{\dag}\eta_{-k}|\Psi_{k}(t)\rangle \\
        & = |u_{k}(0)|^{2}\cos^{2}{\theta_{k}^{f}} + |v_{k}(0)|^{2}\sin^{2}{\theta_{k}^{f}} + i[v_{k}^{*}(0)u_{k}(0)-u_{k}^{*}(0)v_{k}(0)]\sin{\theta_{k}^{f}}\cos{\theta_{k}^{f}}.
    \end{split}
\end{equation}
These results indicate that the steady state $C_{mn}(\infty)$ only depends on the average single fermion occupation numbers $\eta_{k}^{\dag}\eta_{k}$ and$ \eta_{-k}^{\dag}\eta_{-k}$, while the deviation $\delta C_{mn}(t)$ only involves the average double occupation numbers $\eta_{k}^{\dag}\eta_{-k}^{\dag}$ \cite{cao2024relaxation}. As a result, we obtain
\begin{equation}
    \begin{split}
        C_{mn}(k,\infty) & = [\cos^{2}{\theta_{k}^{f}}\langle\Psi_{k}(t)|\eta_{k}^{\dag}\eta_{k}|\Psi(t)\rangle + \sin^{2}{\theta_{k}^{f}}\langle\Psi_{k}(t)|\eta_{-k}\eta_{-k}^{\dag}|\Psi_{k}(t)\rangle]e^{ik(n-m)} \\
                       & + [\cos^{2}{\theta_{k}^{f}}\langle\Psi_{k}(t)|\eta_{-k}^{\dag}\eta_{-k}|\Psi(t)\rangle + \sin^{2}{\theta_{k}^{f}}\langle\Psi_{k}(t)|\eta_{k}\eta_{k}^{\dag}|\Psi_{k}(t)\rangle]e^{-ik(n-m)} \\
                       & = 2[\cos^{2}{\theta_{k}^{f}}\langle\Psi_{k}(t)|\eta_{k}^{\dag}\eta_{k}|\Psi(t)\rangle + \sin^{2}{\theta_{k}^{f}}\langle\Psi_{k}(t)|\eta_{-k}\eta_{-k}^{\dag}|\Psi_{k}(t)\rangle]\cos{k(n-m)} \\
                       & = [|u_{k}(0)|^{2}\sin^{2}{\theta_{k}^{f}}\cos^{2}{\theta_{k}^{f}} + |v_{k}(0)|^{2}\cos^{4}{\theta_{k}^{f}} + i[u_{k}^{*}(0)v_{k}(0)-v_{k}^{*}(0)u_{k}(0)]\sin{\theta_{k}^{f}}\cos^{3}{\theta_{k}^{f}}  \\
                       & + |u_{k}(0)|^{2}\sin^{2}{\theta_{k}^{f}}\cos^{2}{\theta_{k}^{f}} + |v_{k}(0)|^{2}\sin^{4}{\theta_{k}^{f}} + i[v_{k}^{*}(0)u_{k}(0)-u_{k}^{*}(0)v_{k}(0)]\sin^{3}{\theta_{k}^{f}}\cos{\theta_{k}^{f}}]\cos{k(n-m)}  \\
                       & = \{|v_{k}(0)|^{2}+\frac{1}{2}(|u_{k}(0)|^{2}-|v_{k}(0)|^{2})\sin^{2}{2\theta_{k}^{f}}+\frac{1}{4}i[u_{k}^{*}(0)v_{k}(0)-u_{k}(0)v_{k}^{*}(0)]\sin{4\theta_{k}^{f}}\}\cos{k(n-m)},
    \end{split}
\end{equation}
and
\begin{equation}
    \begin{split}
        \delta C_{mn}(k,t) & = i\sin{\theta_{k}^{f}}\cos{\theta_{k}^{f}}[-\langle\Psi_{k}(t)|\eta_{k}^{\dag}\eta_{-k}^{\dag}|\Psi_{k}(t)\rangle + \langle\Psi_{k}(t)|\eta_{-k}\eta_{k}|\Psi_{k}(t)\rangle]e^{ik(n-m)} \\
                         & + i\sin{\theta_{k}^{f}}\cos{\theta_{k}^{f}}[\langle\Psi_{k}(t)|\eta_{-k}^{\dag}\eta_{k}^{\dag}|\Psi_{k}(t)\rangle - \langle\Psi_{k}(t)|\eta_{k}\eta_{-k}|\Psi_{k}(t)\rangle]e^{-ik(n-m)} \\
                         & = 2i\sin{\theta_{k}^{f}}\cos{\theta_{k}^{f}}[-\langle\Psi_{k}(t)|\eta_{k}^{\dag}\eta_{-k}^{\dag}|\Psi_{k}(t)\rangle + \langle\Psi_{k}(t)|\eta_{-k}\eta_{k}|\Psi_{k}(t)\rangle]\cos{k(n-m)}. % \\
                         %& = 2\sin{2\theta_{k}^{f}}[-X_{k}\sin{(2\varepsilon_{k}^{f}t)}+Y_{k}\cos{(2\varepsilon_{k}^{f}t)}]\cos{k(n-m)}
    \end{split} 
\end{equation}
From
\begin{equation}
    \begin{split}
        \langle\eta_{-k}\eta_{k}\rangle_{t} -\langle\eta_{k}^{\dag}\eta_{-k}^{\dag}\rangle_{t} & = \{u_{k}^{*}(0)v_{k}(0)\cos^{2}{\theta_{k}^{f}} + v_{k}^{*}(0)u_{k}(0)\sin^{2}{\theta_{k}^{f}} + i[|v_{k}(0)|^{2}-|u_{k}(0)|^{2}]\sin{\theta_{k}^{f}}\cos{\theta_{k}^{f}}\}e^{2i\varepsilon_{k}^{f}t} \\
        & - \{u_{k}^{*}(0)v_{k}(0)\sin^{2}{\theta_{k}^{f}} + v_{k}^{*}(0)u_{k}(0)\cos^{2}{\theta_{k}^{f}} + i[|u_{k}(0)|^{2}-|v_{k}(0)|^{2}]\sin{\theta_{k}^{f}}\cos{\theta_{k}^{f}}\}e^{-2i\varepsilon_{k}^{f}t} \\
        & = P_{k}e^{2i\varepsilon_{k}^{f}t} - Q_{k}e^{-2i\varepsilon_{k}^{f}t} = (P_{k}-Q_{k})\cos{(2\varepsilon_{k}^{f}t)} + i(P_{k}+Q_{k})\sin{(2\varepsilon_{k}^{f}t)} \\
        & = [(u_{k}^{*}v_{k} - v_{k}^{*}u_{k})\cos{2\theta_{k}^{f}} + i(|v_{k}|^{2}-|u_{k}|^{2})\sin{2\theta_{k}^{f}}]\cos{(2\varepsilon_{k}^{f}t)} + i(u_{k}^{*}v_{k}+v_{k}^{*}u_{k})\sin{(2\varepsilon_{k}^{f}t)},
    \end{split}
\end{equation}
we have
\begin{equation*}
    \begin{split}
        \delta C_{mn}(k,t) & = i\sin{2\theta_{k}^{f}} \{[(u_{k}^{*}v_{k} - v_{k}^{*}u_{k})\cos{2\theta_{k}^{f}} + i(|v_{k}|^{2}-|u_{k}|^{2})\sin{2\theta_{k}^{f}}]\cos{(2\varepsilon_{k}^{f}t)} + i(u_{k}^{*}v_{k}+v_{k}^{*}u_{k})\sin{(2\varepsilon_{k}^{f}t)}\} \\
        & = [i(u_{k}^{*}v_{k}-v_{k}^{*}u_{k})\sin{2\theta_{k}^{f}}\cos{2\theta_{k}^{f}} + (|u_{k}|^{2}-|v_{k}|^{2})\sin^{2}{2\theta_{k}^{f}}]\cos{(2\varepsilon_{k}^{f}t)} -(u_{k}^{*}v_{k}+v_{k}^{*}u_{k})\sin{2\theta_{k}^{f}}\sin{(2\varepsilon_{k}^{f}t)}.
    \end{split}
\end{equation*}
By defining $u_{k} = p_{k}+iq_{k}$ and $v_{k}=r_{k}+is_{k}$, we have
\begin{equation}
    u_{k}^{*}v_{k}-v_{k}^{*}u_{k} = (p_{k}-iq_{k})(r_{k}+is_{k})-(r_{k}-is_{k})(p_{k}+iq_{k}) = 2i(p_{k}s_{k}-q_{k}r_{k}),
\end{equation}
and 
\begin{equation}
    u_{k}^{*}v_{k}+v_{k}^{*}u_{k} = 2(p_{k}r_{k}+q_{k}s_{k}).
\end{equation}
Finally, we obtain the expression of $\delta C_{mn}(k,t)$ as
\begin{equation}
    \delta C_{mn}(k,t) = [(q_{k}r_{k}-p_{k}s_{k})\sin{4\theta_{k}^{f}}+(p_{k}^{2}+q_{k}^{2}-r_{k}^{2}-s_{k}^{2})\sin^{2}{2\theta_{k}^{f}}]\cos{(2\varepsilon_{k}^{f}t)} - 2(p_{k}r_{k}+q_{k}s_{k})\sin{2\theta_{k}^{f}}\sin{(2\varepsilon_{k}^{f}t)}.
\end{equation}
Similarly, the steady value is obtained as 
\begin{equation}
    C_{mn}(k,\infty) = [r_{k}^{2}+s_{k}^{2}+\frac{1}{2}(p_{k}^{2}+q_{k}^{2}-r_{k}^{2}-s_{k}^{2})\sin^{2}{2\theta_{k}^{f}}-\frac{1}{2}(p_{k}s_{k}-q_{k}r_{k})\sin{4\theta_{k}^{f}}]\cos{k(n-m)}.
\end{equation}
   
\end{widetext}

\bibliography{nonequilibirum}

\providecommand{\noopsort}[1]{}\providecommand{\singleletter}[1]{#1}%
\begin{thebibliography}{82}%
\makeatletter
\providecommand \@ifxundefined [1]{%
 \@ifx{#1\undefined}
}%
\providecommand \@ifnum [1]{%
 \ifnum #1\expandafter \@firstoftwo
 \else \expandafter \@secondoftwo
 \fi
}%
\providecommand \@ifx [1]{%
 \ifx #1\expandafter \@firstoftwo
 \else \expandafter \@secondoftwo
 \fi
}%
\providecommand \natexlab [1]{#1}%
\providecommand \enquote  [1]{``#1''}%
\providecommand \bibnamefont  [1]{#1}%
\providecommand \bibfnamefont [1]{#1}%
\providecommand \citenamefont [1]{#1}%
\providecommand \href@noop [0]{\@secondoftwo}%
\providecommand \href [0]{\begingroup \@sanitize@url \@href}%
\providecommand \@href[1]{\@@startlink{#1}\@@href}%
\providecommand \@@href[1]{\endgroup#1\@@endlink}%
\providecommand \@sanitize@url [0]{\catcode `\\12\catcode `\$12\catcode
  `\&12\catcode `\#12\catcode `\^12\catcode `\_12\catcode `\%12\relax}%
\providecommand \@@startlink[1]{}%
\providecommand \@@endlink[0]{}%
\providecommand \url  [0]{\begingroup\@sanitize@url \@url }%
\providecommand \@url [1]{\endgroup\@href {#1}{\urlprefix }}%
\providecommand \urlprefix  [0]{URL }%
\providecommand \Eprint [0]{\href }%
\providecommand \doibase [0]{http://dx.doi.org/}%
\providecommand \selectlanguage [0]{\@gobble}%
\providecommand \bibinfo  [0]{\@secondoftwo}%
\providecommand \bibfield  [0]{\@secondoftwo}%
\providecommand \translation [1]{[#1]}%
\providecommand \BibitemOpen [0]{}%
\providecommand \bibitemStop [0]{}%
\providecommand \bibitemNoStop [0]{.\EOS\space}%
\providecommand \EOS [0]{\spacefactor3000\relax}%
\providecommand \BibitemShut  [1]{\csname bibitem#1\endcsname}%
\let\auto@bib@innerbib\@empty
%</preamble>
\bibitem [{\citenamefont {Sondhi}\ \emph {et~al.}(1997)\citenamefont {Sondhi},
  \citenamefont {Girvin}, \citenamefont {Carini},\ and\ \citenamefont
  {Shahar}}]{Sondhi1999rmp}%
  \BibitemOpen
  \bibfield  {author} {\bibinfo {author} {\bibfnamefont {S.~L.}\ \bibnamefont
  {Sondhi}}, \bibinfo {author} {\bibfnamefont {S.~M.}\ \bibnamefont {Girvin}},
  \bibinfo {author} {\bibfnamefont {J.~P.}\ \bibnamefont {Carini}}, \ and\
  \bibinfo {author} {\bibfnamefont {D.}~\bibnamefont {Shahar}},\ }\href
  {\doibase 10.1103/RevModPhys.69.315} {\bibfield  {journal} {\bibinfo
  {journal} {Rev. Mod. Phys.}\ }\textbf {\bibinfo {volume} {69}},\ \bibinfo
  {pages} {315} (\bibinfo {year} {1997})}\BibitemShut {NoStop}%
\bibitem [{\citenamefont {Sachdev}(2011)}]{Sachdev2011}%
  \BibitemOpen
  \bibfield  {author} {\bibinfo {author} {\bibfnamefont {S.}~\bibnamefont
  {Sachdev}},\ }\href {\doibase 10.1017/CBO9780511973765} {\emph {\bibinfo
  {title} {Quantum Phase Transitions}}},\ \bibinfo {edition} {2nd}\ ed.\
  (\bibinfo  {publisher} {Cambridge University Press},\ \bibinfo {year}
  {2011})\BibitemShut {NoStop}%
\bibitem [{\citenamefont {Calabrese}\ and\ \citenamefont
  {Cardy}(2006)}]{Calabrese2006prl}%
  \BibitemOpen
  \bibfield  {author} {\bibinfo {author} {\bibfnamefont {P.}~\bibnamefont
  {Calabrese}}\ and\ \bibinfo {author} {\bibfnamefont {J.}~\bibnamefont
  {Cardy}},\ }\href {\doibase 10.1103/PhysRevLett.96.136801} {\bibfield
  {journal} {\bibinfo  {journal} {Phys. Rev. Lett.}\ }\textbf {\bibinfo
  {volume} {96}},\ \bibinfo {pages} {136801} (\bibinfo {year}
  {2006})}\BibitemShut {NoStop}%
\bibitem [{\citenamefont {Mondal}\ \emph {et~al.}(2010)\citenamefont {Mondal},
  \citenamefont {Sen},\ and\ \citenamefont {Sengupta}}]{Mondal2010}%
  \BibitemOpen
  \bibfield  {author} {\bibinfo {author} {\bibfnamefont {S.}~\bibnamefont
  {Mondal}}, \bibinfo {author} {\bibfnamefont {D.}~\bibnamefont {Sen}}, \ and\
  \bibinfo {author} {\bibfnamefont {K.}~\bibnamefont {Sengupta}},\ }\href
  {\doibase 10.1007/978-3-642-11470-0} {\emph {\bibinfo {title} {Quantum
  Quenching, Annealing and Computation}}}\ (\bibinfo  {publisher} {Springer
  Berlin Heidelberg},\ \bibinfo {address} {Berlin, Heidelberg},\ \bibinfo
  {year} {2010})\BibitemShut {NoStop}%
\bibitem [{\citenamefont {Dziarmaga}(2010)}]{Dziarmaga2010advp}%
  \BibitemOpen
  \bibfield  {author} {\bibinfo {author} {\bibfnamefont {J.}~\bibnamefont
  {Dziarmaga}},\ }\href {\doibase 10.1080/00018732.2010.514702} {\bibfield
  {journal} {\bibinfo  {journal} {Advances in Physics}\ }\textbf {\bibinfo
  {volume} {59}},\ \bibinfo {pages} {1063} (\bibinfo {year}
  {2010})}\BibitemShut {NoStop}%
\bibitem [{\citenamefont {Polkovnikov}\ \emph {et~al.}(2011)\citenamefont
  {Polkovnikov}, \citenamefont {Sengupta}, \citenamefont {Silva},\ and\
  \citenamefont {Vengalattore}}]{Polkovnikov2011rmp}%
  \BibitemOpen
  \bibfield  {author} {\bibinfo {author} {\bibfnamefont {A.}~\bibnamefont
  {Polkovnikov}}, \bibinfo {author} {\bibfnamefont {K.}~\bibnamefont
  {Sengupta}}, \bibinfo {author} {\bibfnamefont {A.}~\bibnamefont {Silva}}, \
  and\ \bibinfo {author} {\bibfnamefont {M.}~\bibnamefont {Vengalattore}},\
  }\href {\doibase 10.1103/RevModPhys.83.863} {\bibfield  {journal} {\bibinfo
  {journal} {Rev. Mod. Phys.}\ }\textbf {\bibinfo {volume} {83}},\ \bibinfo
  {pages} {863} (\bibinfo {year} {2011})}\BibitemShut {NoStop}%
\bibitem [{\citenamefont {Heyl}\ \emph {et~al.}(2013)\citenamefont {Heyl},
  \citenamefont {Polkovnikov},\ and\ \citenamefont {Kehrein}}]{Heyl2013prl}%
  \BibitemOpen
  \bibfield  {author} {\bibinfo {author} {\bibfnamefont {M.}~\bibnamefont
  {Heyl}}, \bibinfo {author} {\bibfnamefont {A.}~\bibnamefont {Polkovnikov}}, \
  and\ \bibinfo {author} {\bibfnamefont {S.}~\bibnamefont {Kehrein}},\ }\href
  {\doibase 10.1103/PhysRevLett.110.135704} {\bibfield  {journal} {\bibinfo
  {journal} {Phys. Rev. Lett.}\ }\textbf {\bibinfo {volume} {110}},\ \bibinfo
  {pages} {135704} (\bibinfo {year} {2013})}\BibitemShut {NoStop}%
\bibitem [{\citenamefont {Zvyagin}(2016)}]{Zvyagin2016ltp}%
  \BibitemOpen
  \bibfield  {author} {\bibinfo {author} {\bibfnamefont {A.~A.}\ \bibnamefont
  {Zvyagin}},\ }\href {\doibase 10.1063/1.4969869} {\bibfield  {journal}
  {\bibinfo  {journal} {Low Temperature Physics}\ }\textbf {\bibinfo {volume}
  {42}},\ \bibinfo {pages} {971} (\bibinfo {year} {2016})}\BibitemShut
  {NoStop}%
\bibitem [{\citenamefont {Heyl}(2018)}]{Heyl2018rpp}%
  \BibitemOpen
  \bibfield  {author} {\bibinfo {author} {\bibfnamefont {M.}~\bibnamefont
  {Heyl}},\ }\href {\doibase 10.1088/1361-6633/aaaf9a} {\bibfield  {journal}
  {\bibinfo  {journal} {Reports on Progress in Physics}\ }\textbf {\bibinfo
  {volume} {81}},\ \bibinfo {pages} {054001} (\bibinfo {year}
  {2018})}\BibitemShut {NoStop}%
\bibitem [{\citenamefont {Heyl}(2019)}]{Heyl2019el}%
  \BibitemOpen
  \bibfield  {author} {\bibinfo {author} {\bibfnamefont {M.}~\bibnamefont
  {Heyl}},\ }\href {\doibase 10.1209/0295-5075/125/26001} {\bibfield  {journal}
  {\bibinfo  {journal} {Europhysics Letters}\ }\textbf {\bibinfo {volume}
  {125}},\ \bibinfo {pages} {26001} (\bibinfo {year} {2019})}\BibitemShut
  {NoStop}%
\bibitem [{\citenamefont {Prosen}\ and\ \citenamefont
  {Pi\ifmmode~\check{z}\else \v{z}\fi{}orn}(2008)}]{Prosen2008prl}%
  \BibitemOpen
  \bibfield  {author} {\bibinfo {author} {\bibfnamefont {T.~c.~v.}\
  \bibnamefont {Prosen}}\ and\ \bibinfo {author} {\bibfnamefont
  {I.}~\bibnamefont {Pi\ifmmode~\check{z}\else \v{z}\fi{}orn}},\ }\href
  {\doibase 10.1103/PhysRevLett.101.105701} {\bibfield  {journal} {\bibinfo
  {journal} {Phys. Rev. Lett.}\ }\textbf {\bibinfo {volume} {101}},\ \bibinfo
  {pages} {105701} (\bibinfo {year} {2008})}\BibitemShut {NoStop}%
\bibitem [{\citenamefont {Eckstein}\ \emph {et~al.}(2009)\citenamefont
  {Eckstein}, \citenamefont {Kollar},\ and\ \citenamefont
  {Werner}}]{Eckstein2009prl}%
  \BibitemOpen
  \bibfield  {author} {\bibinfo {author} {\bibfnamefont {M.}~\bibnamefont
  {Eckstein}}, \bibinfo {author} {\bibfnamefont {M.}~\bibnamefont {Kollar}}, \
  and\ \bibinfo {author} {\bibfnamefont {P.}~\bibnamefont {Werner}},\ }\href
  {\doibase 10.1103/PhysRevLett.103.056403} {\bibfield  {journal} {\bibinfo
  {journal} {Phys. Rev. Lett.}\ }\textbf {\bibinfo {volume} {103}},\ \bibinfo
  {pages} {056403} (\bibinfo {year} {2009})}\BibitemShut {NoStop}%
\bibitem [{\citenamefont {Barmettler}\ \emph {et~al.}(2009)\citenamefont
  {Barmettler}, \citenamefont {Punk}, \citenamefont {Gritsev}, \citenamefont
  {Demler},\ and\ \citenamefont {Altman}}]{Barmettler2009prl}%
  \BibitemOpen
  \bibfield  {author} {\bibinfo {author} {\bibfnamefont {P.}~\bibnamefont
  {Barmettler}}, \bibinfo {author} {\bibfnamefont {M.}~\bibnamefont {Punk}},
  \bibinfo {author} {\bibfnamefont {V.}~\bibnamefont {Gritsev}}, \bibinfo
  {author} {\bibfnamefont {E.}~\bibnamefont {Demler}}, \ and\ \bibinfo {author}
  {\bibfnamefont {E.}~\bibnamefont {Altman}},\ }\href {\doibase
  10.1103/PhysRevLett.102.130603} {\bibfield  {journal} {\bibinfo  {journal}
  {Phys. Rev. Lett.}\ }\textbf {\bibinfo {volume} {102}},\ \bibinfo {pages}
  {130603} (\bibinfo {year} {2009})}\BibitemShut {NoStop}%
\bibitem [{\citenamefont {Barmettler}\ \emph {et~al.}(2010)\citenamefont
  {Barmettler}, \citenamefont {Punk}, \citenamefont {Gritsev}, \citenamefont
  {Demler},\ and\ \citenamefont {Altman}}]{Barmettler2010njp}%
  \BibitemOpen
  \bibfield  {author} {\bibinfo {author} {\bibfnamefont {P.}~\bibnamefont
  {Barmettler}}, \bibinfo {author} {\bibfnamefont {M.}~\bibnamefont {Punk}},
  \bibinfo {author} {\bibfnamefont {V.}~\bibnamefont {Gritsev}}, \bibinfo
  {author} {\bibfnamefont {E.}~\bibnamefont {Demler}}, \ and\ \bibinfo {author}
  {\bibfnamefont {E.}~\bibnamefont {Altman}},\ }\href {\doibase
  10.1088/1367-2630/12/5/055017} {\bibfield  {journal} {\bibinfo  {journal}
  {New Journal of Physics}\ }\textbf {\bibinfo {volume} {12}},\ \bibinfo
  {pages} {055017} (\bibinfo {year} {2010})}\BibitemShut {NoStop}%
\bibitem [{\citenamefont {Marino}\ and\ \citenamefont
  {Silva}(2012)}]{Marino2012prb}%
  \BibitemOpen
  \bibfield  {author} {\bibinfo {author} {\bibfnamefont {J.}~\bibnamefont
  {Marino}}\ and\ \bibinfo {author} {\bibfnamefont {A.}~\bibnamefont {Silva}},\
  }\href {\doibase 10.1103/PhysRevB.86.060408} {\bibfield  {journal} {\bibinfo
  {journal} {Phys. Rev. B}\ }\textbf {\bibinfo {volume} {86}},\ \bibinfo
  {pages} {060408} (\bibinfo {year} {2012})}\BibitemShut {NoStop}%
\bibitem [{\citenamefont {Greiner}\ \emph {et~al.}(2002)\citenamefont
  {Greiner}, \citenamefont {Mandel}, \citenamefont {Esslinger}, \citenamefont
  {Hänsch},\ and\ \citenamefont {Bloch}}]{Greiner2002nature}%
  \BibitemOpen
  \bibfield  {author} {\bibinfo {author} {\bibfnamefont {M.}~\bibnamefont
  {Greiner}}, \bibinfo {author} {\bibfnamefont {O.}~\bibnamefont {Mandel}},
  \bibinfo {author} {\bibfnamefont {T.}~\bibnamefont {Esslinger}}, \bibinfo
  {author} {\bibfnamefont {T.~W.}\ \bibnamefont {Hänsch}}, \ and\ \bibinfo
  {author} {\bibfnamefont {I.}~\bibnamefont {Bloch}},\ }\href {\doibase
  10.1038/415039a} {\bibfield  {journal} {\bibinfo  {journal} {Nature}\
  }\textbf {\bibinfo {volume} {415}},\ \bibinfo {pages} {39} (\bibinfo {year}
  {2002})}\BibitemShut {NoStop}%
\bibitem [{\citenamefont {Kinoshita}\ \emph {et~al.}(2006)\citenamefont
  {Kinoshita}, \citenamefont {Wenger},\ and\ \citenamefont
  {Weiss}}]{Kinoshita2006nature}%
  \BibitemOpen
  \bibfield  {author} {\bibinfo {author} {\bibfnamefont {T.}~\bibnamefont
  {Kinoshita}}, \bibinfo {author} {\bibfnamefont {T.}~\bibnamefont {Wenger}}, \
  and\ \bibinfo {author} {\bibfnamefont {D.~S.}\ \bibnamefont {Weiss}},\ }\href
  {\doibase 10.1038/nature04693} {\bibfield  {journal} {\bibinfo  {journal}
  {Nature}\ }\textbf {\bibinfo {volume} {440}},\ \bibinfo {pages} {900}
  (\bibinfo {year} {2006})}\BibitemShut {NoStop}%
\bibitem [{\citenamefont {Bloch}\ \emph {et~al.}(2008)\citenamefont {Bloch},
  \citenamefont {Dalibard},\ and\ \citenamefont {Zwerger}}]{Bloch2008rmp}%
  \BibitemOpen
  \bibfield  {author} {\bibinfo {author} {\bibfnamefont {I.}~\bibnamefont
  {Bloch}}, \bibinfo {author} {\bibfnamefont {J.}~\bibnamefont {Dalibard}}, \
  and\ \bibinfo {author} {\bibfnamefont {W.}~\bibnamefont {Zwerger}},\ }\href
  {\doibase 10.1103/RevModPhys.80.885} {\bibfield  {journal} {\bibinfo
  {journal} {Rev. Mod. Phys.}\ }\textbf {\bibinfo {volume} {80}},\ \bibinfo
  {pages} {885} (\bibinfo {year} {2008})}\BibitemShut {NoStop}%
\bibitem [{\citenamefont {Georgescu}\ \emph {et~al.}(2014)\citenamefont
  {Georgescu}, \citenamefont {Ashhab},\ and\ \citenamefont
  {Nori}}]{Georgescu2014rmp}%
  \BibitemOpen
  \bibfield  {author} {\bibinfo {author} {\bibfnamefont {I.~M.}\ \bibnamefont
  {Georgescu}}, \bibinfo {author} {\bibfnamefont {S.}~\bibnamefont {Ashhab}}, \
  and\ \bibinfo {author} {\bibfnamefont {F.}~\bibnamefont {Nori}},\ }\href
  {\doibase 10.1103/RevModPhys.86.153} {\bibfield  {journal} {\bibinfo
  {journal} {Rev. Mod. Phys.}\ }\textbf {\bibinfo {volume} {86}},\ \bibinfo
  {pages} {153} (\bibinfo {year} {2014})}\BibitemShut {NoStop}%
\bibitem [{\citenamefont {Calabrese}\ and\ \citenamefont
  {Cardy}(2007)}]{Calabrese2007jsmte}%
  \BibitemOpen
  \bibfield  {author} {\bibinfo {author} {\bibfnamefont {P.}~\bibnamefont
  {Calabrese}}\ and\ \bibinfo {author} {\bibfnamefont {J.}~\bibnamefont
  {Cardy}},\ }\href {\doibase 10.1088/1742-5468/2007/06/P06008} {\bibfield
  {journal} {\bibinfo  {journal} {Journal of Statistical Mechanics: Theory and
  Experiment}\ }\textbf {\bibinfo {volume} {2007}},\ \bibinfo {pages} {P06008}
  (\bibinfo {year} {2007})}\BibitemShut {NoStop}%
\bibitem [{\citenamefont {Iucci}\ and\ \citenamefont
  {Cazalilla}(2009)}]{Iucci2009pra}%
  \BibitemOpen
  \bibfield  {author} {\bibinfo {author} {\bibfnamefont {A.}~\bibnamefont
  {Iucci}}\ and\ \bibinfo {author} {\bibfnamefont {M.~A.}\ \bibnamefont
  {Cazalilla}},\ }\href {\doibase 10.1103/PhysRevA.80.063619} {\bibfield
  {journal} {\bibinfo  {journal} {Phys. Rev. A}\ }\textbf {\bibinfo {volume}
  {80}},\ \bibinfo {pages} {063619} (\bibinfo {year} {2009})}\BibitemShut
  {NoStop}%
\bibitem [{\citenamefont {Manmana}\ \emph {et~al.}(2009)\citenamefont
  {Manmana}, \citenamefont {Wessel}, \citenamefont {Noack},\ and\ \citenamefont
  {Muramatsu}}]{Manmana2009prb}%
  \BibitemOpen
  \bibfield  {author} {\bibinfo {author} {\bibfnamefont {S.~R.}\ \bibnamefont
  {Manmana}}, \bibinfo {author} {\bibfnamefont {S.}~\bibnamefont {Wessel}},
  \bibinfo {author} {\bibfnamefont {R.~M.}\ \bibnamefont {Noack}}, \ and\
  \bibinfo {author} {\bibfnamefont {A.}~\bibnamefont {Muramatsu}},\ }\href
  {\doibase 10.1103/PhysRevB.79.155104} {\bibfield  {journal} {\bibinfo
  {journal} {Phys. Rev. B}\ }\textbf {\bibinfo {volume} {79}},\ \bibinfo
  {pages} {155104} (\bibinfo {year} {2009})}\BibitemShut {NoStop}%
\bibitem [{\citenamefont {Brandino}\ \emph {et~al.}(2012)\citenamefont
  {Brandino}, \citenamefont {De~Luca}, \citenamefont {Konik},\ and\
  \citenamefont {Mussardo}}]{Brandino2012prb}%
  \BibitemOpen
  \bibfield  {author} {\bibinfo {author} {\bibfnamefont {G.~P.}\ \bibnamefont
  {Brandino}}, \bibinfo {author} {\bibfnamefont {A.}~\bibnamefont {De~Luca}},
  \bibinfo {author} {\bibfnamefont {R.~M.}\ \bibnamefont {Konik}}, \ and\
  \bibinfo {author} {\bibfnamefont {G.}~\bibnamefont {Mussardo}},\ }\href
  {\doibase 10.1103/PhysRevB.85.214435} {\bibfield  {journal} {\bibinfo
  {journal} {Phys. Rev. B}\ }\textbf {\bibinfo {volume} {85}},\ \bibinfo
  {pages} {214435} (\bibinfo {year} {2012})}\BibitemShut {NoStop}%
\bibitem [{\citenamefont {Rigol}(2014)}]{Rigol2014prl}%
  \BibitemOpen
  \bibfield  {author} {\bibinfo {author} {\bibfnamefont {M.}~\bibnamefont
  {Rigol}},\ }\href {\doibase 10.1103/PhysRevLett.112.170601} {\bibfield
  {journal} {\bibinfo  {journal} {Phys. Rev. Lett.}\ }\textbf {\bibinfo
  {volume} {112}},\ \bibinfo {pages} {170601} (\bibinfo {year}
  {2014})}\BibitemShut {NoStop}%
\bibitem [{\citenamefont {Collura}\ and\ \citenamefont
  {Karevski}(2014)}]{Collura2014prb}%
  \BibitemOpen
  \bibfield  {author} {\bibinfo {author} {\bibfnamefont {M.}~\bibnamefont
  {Collura}}\ and\ \bibinfo {author} {\bibfnamefont {D.}~\bibnamefont
  {Karevski}},\ }\href {\doibase 10.1103/PhysRevB.89.214308} {\bibfield
  {journal} {\bibinfo  {journal} {Phys. Rev. B}\ }\textbf {\bibinfo {volume}
  {89}},\ \bibinfo {pages} {214308} (\bibinfo {year} {2014})}\BibitemShut
  {NoStop}%
\bibitem [{\citenamefont {Lindner}\ \emph {et~al.}(2011)\citenamefont
  {Lindner}, \citenamefont {Refael},\ and\ \citenamefont
  {Galitski}}]{Lindner2011nphys}%
  \BibitemOpen
  \bibfield  {author} {\bibinfo {author} {\bibfnamefont {N.~H.}\ \bibnamefont
  {Lindner}}, \bibinfo {author} {\bibfnamefont {G.}~\bibnamefont {Refael}}, \
  and\ \bibinfo {author} {\bibfnamefont {V.}~\bibnamefont {Galitski}},\ }\href
  {\doibase 10.1038/nphys1926} {\bibfield  {journal} {\bibinfo  {journal}
  {Nature Physics}\ }\textbf {\bibinfo {volume} {7}},\ \bibinfo {pages} {490}
  (\bibinfo {year} {2011})}\BibitemShut {NoStop}%
\bibitem [{\citenamefont {Sacha}\ and\ \citenamefont
  {Zakrzewski}(2017)}]{Sacha2018rpp}%
  \BibitemOpen
  \bibfield  {author} {\bibinfo {author} {\bibfnamefont {K.}~\bibnamefont
  {Sacha}}\ and\ \bibinfo {author} {\bibfnamefont {J.}~\bibnamefont
  {Zakrzewski}},\ }\href {\doibase 10.1088/1361-6633/aa8b38} {\bibfield
  {journal} {\bibinfo  {journal} {Reports on Progress in Physics}\ }\textbf
  {\bibinfo {volume} {81}},\ \bibinfo {pages} {016401} (\bibinfo {year}
  {2017})}\BibitemShut {NoStop}%
\bibitem [{\citenamefont {Kurzyna}\ and\ \citenamefont
  {Kwapi\ifmmode~\acute{n}\else \'{n}\fi{}ski}(2020)}]{Kurzyna2020prb}%
  \BibitemOpen
  \bibfield  {author} {\bibinfo {author} {\bibfnamefont {M.}~\bibnamefont
  {Kurzyna}}\ and\ \bibinfo {author} {\bibfnamefont {T.}~\bibnamefont
  {Kwapi\ifmmode~\acute{n}\else \'{n}\fi{}ski}},\ }\href {\doibase
  10.1103/PhysRevB.102.195429} {\bibfield  {journal} {\bibinfo  {journal}
  {Phys. Rev. B}\ }\textbf {\bibinfo {volume} {102}},\ \bibinfo {pages}
  {195429} (\bibinfo {year} {2020})}\BibitemShut {NoStop}%
\bibitem [{\citenamefont {Andraschko}\ and\ \citenamefont
  {Sirker}(2014)}]{Andraschko2014prb}%
  \BibitemOpen
  \bibfield  {author} {\bibinfo {author} {\bibfnamefont {F.}~\bibnamefont
  {Andraschko}}\ and\ \bibinfo {author} {\bibfnamefont {J.}~\bibnamefont
  {Sirker}},\ }\href {\doibase 10.1103/PhysRevB.89.125120} {\bibfield
  {journal} {\bibinfo  {journal} {Phys. Rev. B}\ }\textbf {\bibinfo {volume}
  {89}},\ \bibinfo {pages} {125120} (\bibinfo {year} {2014})}\BibitemShut
  {NoStop}%
\bibitem [{\citenamefont {Hickey}\ \emph {et~al.}(2014)\citenamefont {Hickey},
  \citenamefont {Genway},\ and\ \citenamefont {Garrahan}}]{Hickey2014prb}%
  \BibitemOpen
  \bibfield  {author} {\bibinfo {author} {\bibfnamefont {J.~M.}\ \bibnamefont
  {Hickey}}, \bibinfo {author} {\bibfnamefont {S.}~\bibnamefont {Genway}}, \
  and\ \bibinfo {author} {\bibfnamefont {J.~P.}\ \bibnamefont {Garrahan}},\
  }\href {\doibase 10.1103/PhysRevB.89.054301} {\bibfield  {journal} {\bibinfo
  {journal} {Phys. Rev. B}\ }\textbf {\bibinfo {volume} {89}},\ \bibinfo
  {pages} {054301} (\bibinfo {year} {2014})}\BibitemShut {NoStop}%
\bibitem [{\citenamefont {Vajna}\ and\ \citenamefont
  {D\'ora}(2014)}]{Vajna2014prb}%
  \BibitemOpen
  \bibfield  {author} {\bibinfo {author} {\bibfnamefont {S.}~\bibnamefont
  {Vajna}}\ and\ \bibinfo {author} {\bibfnamefont {B.}~\bibnamefont {D\'ora}},\
  }\href {\doibase 10.1103/PhysRevB.89.161105} {\bibfield  {journal} {\bibinfo
  {journal} {Phys. Rev. B}\ }\textbf {\bibinfo {volume} {89}},\ \bibinfo
  {pages} {161105} (\bibinfo {year} {2014})}\BibitemShut {NoStop}%
\bibitem [{\citenamefont {Schmitt}\ and\ \citenamefont
  {Kehrein}(2015)}]{Schmitt2015prb}%
  \BibitemOpen
  \bibfield  {author} {\bibinfo {author} {\bibfnamefont {M.}~\bibnamefont
  {Schmitt}}\ and\ \bibinfo {author} {\bibfnamefont {S.}~\bibnamefont
  {Kehrein}},\ }\href {\doibase 10.1103/PhysRevB.92.075114} {\bibfield
  {journal} {\bibinfo  {journal} {Phys. Rev. B}\ }\textbf {\bibinfo {volume}
  {92}},\ \bibinfo {pages} {075114} (\bibinfo {year} {2015})}\BibitemShut
  {NoStop}%
\bibitem [{\citenamefont {Vajna}\ and\ \citenamefont
  {D\'ora}(2015)}]{Vajna2015prb}%
  \BibitemOpen
  \bibfield  {author} {\bibinfo {author} {\bibfnamefont {S.}~\bibnamefont
  {Vajna}}\ and\ \bibinfo {author} {\bibfnamefont {B.}~\bibnamefont {D\'ora}},\
  }\href {\doibase 10.1103/PhysRevB.91.155127} {\bibfield  {journal} {\bibinfo
  {journal} {Phys. Rev. B}\ }\textbf {\bibinfo {volume} {91}},\ \bibinfo
  {pages} {155127} (\bibinfo {year} {2015})}\BibitemShut {NoStop}%
\bibitem [{\citenamefont {Divakaran}\ \emph {et~al.}(2016)\citenamefont
  {Divakaran}, \citenamefont {Sharma},\ and\ \citenamefont
  {Dutta}}]{Divakaran2016pre}%
  \BibitemOpen
  \bibfield  {author} {\bibinfo {author} {\bibfnamefont {U.}~\bibnamefont
  {Divakaran}}, \bibinfo {author} {\bibfnamefont {S.}~\bibnamefont {Sharma}}, \
  and\ \bibinfo {author} {\bibfnamefont {A.}~\bibnamefont {Dutta}},\ }\href
  {\doibase 10.1103/PhysRevE.93.052133} {\bibfield  {journal} {\bibinfo
  {journal} {Phys. Rev. E}\ }\textbf {\bibinfo {volume} {93}},\ \bibinfo
  {pages} {052133} (\bibinfo {year} {2016})}\BibitemShut {NoStop}%
\bibitem [{\citenamefont {Bhattacharya}\ \emph {et~al.}(2017)\citenamefont
  {Bhattacharya}, \citenamefont {Bandyopadhyay},\ and\ \citenamefont
  {Dutta}}]{Bhattacharya2017prb}%
  \BibitemOpen
  \bibfield  {author} {\bibinfo {author} {\bibfnamefont {U.}~\bibnamefont
  {Bhattacharya}}, \bibinfo {author} {\bibfnamefont {S.}~\bibnamefont
  {Bandyopadhyay}}, \ and\ \bibinfo {author} {\bibfnamefont {A.}~\bibnamefont
  {Dutta}},\ }\href {\doibase 10.1103/PhysRevB.96.180303} {\bibfield  {journal}
  {\bibinfo  {journal} {Phys. Rev. B}\ }\textbf {\bibinfo {volume} {96}},\
  \bibinfo {pages} {180303} (\bibinfo {year} {2017})}\BibitemShut {NoStop}%
\bibitem [{\citenamefont {Kosior}\ and\ \citenamefont
  {Sacha}(2018)}]{Kosior2018pra}%
  \BibitemOpen
  \bibfield  {author} {\bibinfo {author} {\bibfnamefont {A.}~\bibnamefont
  {Kosior}}\ and\ \bibinfo {author} {\bibfnamefont {K.}~\bibnamefont {Sacha}},\
  }\href {\doibase 10.1103/PhysRevA.97.053621} {\bibfield  {journal} {\bibinfo
  {journal} {Phys. Rev. A}\ }\textbf {\bibinfo {volume} {97}},\ \bibinfo
  {pages} {053621} (\bibinfo {year} {2018})}\BibitemShut {NoStop}%
\bibitem [{\citenamefont {Lang}\ \emph
  {et~al.}(2018{\natexlab{a}})\citenamefont {Lang}, \citenamefont {Chen},
  \citenamefont {Hong},\ and\ \citenamefont {Fan}}]{Lang2018prb}%
  \BibitemOpen
  \bibfield  {author} {\bibinfo {author} {\bibfnamefont {H.}~\bibnamefont
  {Lang}}, \bibinfo {author} {\bibfnamefont {Y.}~\bibnamefont {Chen}}, \bibinfo
  {author} {\bibfnamefont {Q.}~\bibnamefont {Hong}}, \ and\ \bibinfo {author}
  {\bibfnamefont {H.}~\bibnamefont {Fan}},\ }\href {\doibase
  10.1103/PhysRevB.98.134310} {\bibfield  {journal} {\bibinfo  {journal} {Phys.
  Rev. B}\ }\textbf {\bibinfo {volume} {98}},\ \bibinfo {pages} {134310}
  (\bibinfo {year} {2018}{\natexlab{a}})}\BibitemShut {NoStop}%
\bibitem [{\citenamefont {Lahiri}\ and\ \citenamefont
  {Bera}(2019{\natexlab{a}})}]{Lahiri2019prb}%
  \BibitemOpen
  \bibfield  {author} {\bibinfo {author} {\bibfnamefont {A.}~\bibnamefont
  {Lahiri}}\ and\ \bibinfo {author} {\bibfnamefont {S.}~\bibnamefont {Bera}},\
  }\href {\doibase 10.1103/PhysRevB.99.174311} {\bibfield  {journal} {\bibinfo
  {journal} {Phys. Rev. B}\ }\textbf {\bibinfo {volume} {99}},\ \bibinfo
  {pages} {174311} (\bibinfo {year} {2019}{\natexlab{a}})}\BibitemShut
  {NoStop}%
\bibitem [{\citenamefont {Liu}\ and\ \citenamefont {Guo}(2019)}]{Liu2019prb}%
  \BibitemOpen
  \bibfield  {author} {\bibinfo {author} {\bibfnamefont {T.}~\bibnamefont
  {Liu}}\ and\ \bibinfo {author} {\bibfnamefont {H.}~\bibnamefont {Guo}},\
  }\href {\doibase 10.1103/PhysRevB.99.104307} {\bibfield  {journal} {\bibinfo
  {journal} {Phys. Rev. B}\ }\textbf {\bibinfo {volume} {99}},\ \bibinfo
  {pages} {104307} (\bibinfo {year} {2019})}\BibitemShut {NoStop}%
\bibitem [{\citenamefont {Cao}\ \emph {et~al.}(2020)\citenamefont {Cao},
  \citenamefont {Li}, \citenamefont {Zhong},\ and\ \citenamefont
  {Tong}}]{Cao2020prb}%
  \BibitemOpen
  \bibfield  {author} {\bibinfo {author} {\bibfnamefont {K.}~\bibnamefont
  {Cao}}, \bibinfo {author} {\bibfnamefont {W.}~\bibnamefont {Li}}, \bibinfo
  {author} {\bibfnamefont {M.}~\bibnamefont {Zhong}}, \ and\ \bibinfo {author}
  {\bibfnamefont {P.}~\bibnamefont {Tong}},\ }\href {\doibase
  10.1103/PhysRevB.102.014207} {\bibfield  {journal} {\bibinfo  {journal}
  {Phys. Rev. B}\ }\textbf {\bibinfo {volume} {102}},\ \bibinfo {pages}
  {014207} (\bibinfo {year} {2020})}\BibitemShut {NoStop}%
\bibitem [{\citenamefont {Cao}\ \emph {et~al.}(2023{\natexlab{a}})\citenamefont
  {Cao}, \citenamefont {Yang}, \citenamefont {Hu},\ and\ \citenamefont
  {Yang}}]{Cao2023prb}%
  \BibitemOpen
  \bibfield  {author} {\bibinfo {author} {\bibfnamefont {K.}~\bibnamefont
  {Cao}}, \bibinfo {author} {\bibfnamefont {S.}~\bibnamefont {Yang}}, \bibinfo
  {author} {\bibfnamefont {Y.}~\bibnamefont {Hu}}, \ and\ \bibinfo {author}
  {\bibfnamefont {G.}~\bibnamefont {Yang}},\ }\href {\doibase
  10.1103/PhysRevB.108.024201} {\bibfield  {journal} {\bibinfo  {journal}
  {Phys. Rev. B}\ }\textbf {\bibinfo {volume} {108}},\ \bibinfo {pages}
  {024201} (\bibinfo {year} {2023}{\natexlab{a}})}\BibitemShut {NoStop}%
\bibitem [{\citenamefont {Kuliashov}\ \emph {et~al.}(2023)\citenamefont
  {Kuliashov}, \citenamefont {Markov},\ and\ \citenamefont
  {Rubtsov}}]{Kuliashov2023prb}%
  \BibitemOpen
  \bibfield  {author} {\bibinfo {author} {\bibfnamefont {O.~N.}\ \bibnamefont
  {Kuliashov}}, \bibinfo {author} {\bibfnamefont {A.~A.}\ \bibnamefont
  {Markov}}, \ and\ \bibinfo {author} {\bibfnamefont {A.~N.}\ \bibnamefont
  {Rubtsov}},\ }\href {\doibase 10.1103/PhysRevB.107.094304} {\bibfield
  {journal} {\bibinfo  {journal} {Phys. Rev. B}\ }\textbf {\bibinfo {volume}
  {107}},\ \bibinfo {pages} {094304} (\bibinfo {year} {2023})}\BibitemShut
  {NoStop}%
\bibitem [{\citenamefont {Sacramento}\ and\ \citenamefont
  {Yu}(2024)}]{Sacramento2024prb}%
  \BibitemOpen
  \bibfield  {author} {\bibinfo {author} {\bibfnamefont {P.~D.}\ \bibnamefont
  {Sacramento}}\ and\ \bibinfo {author} {\bibfnamefont {W.~C.}\ \bibnamefont
  {Yu}},\ }\href {\doibase 10.1103/PhysRevB.109.134301} {\bibfield  {journal}
  {\bibinfo  {journal} {Phys. Rev. B}\ }\textbf {\bibinfo {volume} {109}},\
  \bibinfo {pages} {134301} (\bibinfo {year} {2024})}\BibitemShut {NoStop}%
\bibitem [{\citenamefont {Dziarmaga}(2005)}]{Dziarmaga2005prl}%
  \BibitemOpen
  \bibfield  {author} {\bibinfo {author} {\bibfnamefont {J.}~\bibnamefont
  {Dziarmaga}},\ }\href {\doibase 10.1103/PhysRevLett.95.245701} {\bibfield
  {journal} {\bibinfo  {journal} {Phys. Rev. Lett.}\ }\textbf {\bibinfo
  {volume} {95}},\ \bibinfo {pages} {245701} (\bibinfo {year}
  {2005})}\BibitemShut {NoStop}%
\bibitem [{\citenamefont {De~Grandi}\ \emph {et~al.}(2010)\citenamefont
  {De~Grandi}, \citenamefont {Gritsev},\ and\ \citenamefont
  {Polkovnikov}}]{Grandi2010prb}%
  \BibitemOpen
  \bibfield  {author} {\bibinfo {author} {\bibfnamefont {C.}~\bibnamefont
  {De~Grandi}}, \bibinfo {author} {\bibfnamefont {V.}~\bibnamefont {Gritsev}},
  \ and\ \bibinfo {author} {\bibfnamefont {A.}~\bibnamefont {Polkovnikov}},\
  }\href {\doibase 10.1103/PhysRevB.81.012303} {\bibfield  {journal} {\bibinfo
  {journal} {Phys. Rev. B}\ }\textbf {\bibinfo {volume} {81}},\ \bibinfo
  {pages} {012303} (\bibinfo {year} {2010})}\BibitemShut {NoStop}%
\bibitem [{\citenamefont {Weiss}\ \emph {et~al.}(2018)\citenamefont {Weiss},
  \citenamefont {Gerster}, \citenamefont {Jaschke}, \citenamefont {Silvi},\
  and\ \citenamefont {Montangero}}]{Weiss2018pra}%
  \BibitemOpen
  \bibfield  {author} {\bibinfo {author} {\bibfnamefont {W.}~\bibnamefont
  {Weiss}}, \bibinfo {author} {\bibfnamefont {M.}~\bibnamefont {Gerster}},
  \bibinfo {author} {\bibfnamefont {D.}~\bibnamefont {Jaschke}}, \bibinfo
  {author} {\bibfnamefont {P.}~\bibnamefont {Silvi}}, \ and\ \bibinfo {author}
  {\bibfnamefont {S.}~\bibnamefont {Montangero}},\ }\href {\doibase
  10.1103/PhysRevA.98.063601} {\bibfield  {journal} {\bibinfo  {journal} {Phys.
  Rev. A}\ }\textbf {\bibinfo {volume} {98}},\ \bibinfo {pages} {063601}
  (\bibinfo {year} {2018})}\BibitemShut {NoStop}%
\bibitem [{\citenamefont {Balducci}\ \emph {et~al.}(2023)\citenamefont
  {Balducci}, \citenamefont {Beau}, \citenamefont {Yang}, \citenamefont
  {Gambassi},\ and\ \citenamefont {del Campo}}]{Balducci2023prl}%
  \BibitemOpen
  \bibfield  {author} {\bibinfo {author} {\bibfnamefont {F.}~\bibnamefont
  {Balducci}}, \bibinfo {author} {\bibfnamefont {M.}~\bibnamefont {Beau}},
  \bibinfo {author} {\bibfnamefont {J.}~\bibnamefont {Yang}}, \bibinfo {author}
  {\bibfnamefont {A.}~\bibnamefont {Gambassi}}, \ and\ \bibinfo {author}
  {\bibfnamefont {A.}~\bibnamefont {del Campo}},\ }\href {\doibase
  10.1103/PhysRevLett.131.230401} {\bibfield  {journal} {\bibinfo  {journal}
  {Phys. Rev. Lett.}\ }\textbf {\bibinfo {volume} {131}},\ \bibinfo {pages}
  {230401} (\bibinfo {year} {2023})}\BibitemShut {NoStop}%
\bibitem [{\citenamefont {Sharma}\ \emph {et~al.}(2016)\citenamefont {Sharma},
  \citenamefont {Divakaran}, \citenamefont {Polkovnikov},\ and\ \citenamefont
  {Dutta}}]{Sharma2016prb}%
  \BibitemOpen
  \bibfield  {author} {\bibinfo {author} {\bibfnamefont {S.}~\bibnamefont
  {Sharma}}, \bibinfo {author} {\bibfnamefont {U.}~\bibnamefont {Divakaran}},
  \bibinfo {author} {\bibfnamefont {A.}~\bibnamefont {Polkovnikov}}, \ and\
  \bibinfo {author} {\bibfnamefont {A.}~\bibnamefont {Dutta}},\ }\href
  {\doibase 10.1103/PhysRevB.93.144306} {\bibfield  {journal} {\bibinfo
  {journal} {Phys. Rev. B}\ }\textbf {\bibinfo {volume} {93}},\ \bibinfo
  {pages} {144306} (\bibinfo {year} {2016})}\BibitemShut {NoStop}%
\bibitem [{\citenamefont {Zamani}\ \emph {et~al.}(2024)\citenamefont {Zamani},
  \citenamefont {Naji}, \citenamefont {Jafari},\ and\ \citenamefont
  {Langari}}]{zamani2310}%
  \BibitemOpen
  \bibfield  {author} {\bibinfo {author} {\bibfnamefont {S.}~\bibnamefont
  {Zamani}}, \bibinfo {author} {\bibfnamefont {J.}~\bibnamefont {Naji}},
  \bibinfo {author} {\bibfnamefont {R.}~\bibnamefont {Jafari}}, \ and\ \bibinfo
  {author} {\bibfnamefont {A.}~\bibnamefont {Langari}},\ }\href {\doibase
  10.48550/arXiv.2310.15101} {\enquote {\bibinfo {title} {Scaling and
  universality at ramped quench dynamical quantum phase transition},}\ }
  (\bibinfo {year} {2024}),\ \Eprint {http://arxiv.org/abs/2310.15101}
  {arXiv:2310.15101 [cond-mat.stat-mech]} \BibitemShut {NoStop}%
\bibitem [{\citenamefont {Puskarov}\ and\ \citenamefont
  {Schuricht}(2016)}]{Tatjana2016SciPost}%
  \BibitemOpen
  \bibfield  {author} {\bibinfo {author} {\bibfnamefont {T.}~\bibnamefont
  {Puskarov}}\ and\ \bibinfo {author} {\bibfnamefont {D.}~\bibnamefont
  {Schuricht}},\ }\href {\doibase 10.21468/SciPostPhys.1.1.003} {\bibfield
  {journal} {\bibinfo  {journal} {SciPost Phys.}\ }\textbf {\bibinfo {volume}
  {1}},\ \bibinfo {pages} {003} (\bibinfo {year} {2016})}\BibitemShut {NoStop}%
\bibitem [{\citenamefont {Sen}\ \emph {et~al.}(2016)\citenamefont {Sen},
  \citenamefont {Nandy},\ and\ \citenamefont {Sengupta}}]{Sen2016prb}%
  \BibitemOpen
  \bibfield  {author} {\bibinfo {author} {\bibfnamefont {A.}~\bibnamefont
  {Sen}}, \bibinfo {author} {\bibfnamefont {S.}~\bibnamefont {Nandy}}, \ and\
  \bibinfo {author} {\bibfnamefont {K.}~\bibnamefont {Sengupta}},\ }\href
  {\doibase 10.1103/PhysRevB.94.214301} {\bibfield  {journal} {\bibinfo
  {journal} {Phys. Rev. B}\ }\textbf {\bibinfo {volume} {94}},\ \bibinfo
  {pages} {214301} (\bibinfo {year} {2016})}\BibitemShut {NoStop}%
\bibitem [{\citenamefont {Nandy}\ \emph {et~al.}(2018)\citenamefont {Nandy},
  \citenamefont {Sengupta},\ and\ \citenamefont {Sen}}]{Sourav2018jpa}%
  \BibitemOpen
  \bibfield  {author} {\bibinfo {author} {\bibfnamefont {S.}~\bibnamefont
  {Nandy}}, \bibinfo {author} {\bibfnamefont {K.}~\bibnamefont {Sengupta}}, \
  and\ \bibinfo {author} {\bibfnamefont {A.}~\bibnamefont {Sen}},\ }\href
  {\doibase 10.1088/1751-8121/aaced6} {\bibfield  {journal} {\bibinfo
  {journal} {Journal of Physics A: Mathematical and Theoretical}\ }\textbf
  {\bibinfo {volume} {51}},\ \bibinfo {pages} {334002} (\bibinfo {year}
  {2018})}\BibitemShut {NoStop}%
\bibitem [{\citenamefont {Sarkar}\ and\ \citenamefont
  {Sengupta}(2020)}]{Sarkar2020prb}%
  \BibitemOpen
  \bibfield  {author} {\bibinfo {author} {\bibfnamefont {M.}~\bibnamefont
  {Sarkar}}\ and\ \bibinfo {author} {\bibfnamefont {K.}~\bibnamefont
  {Sengupta}},\ }\href {\doibase 10.1103/PhysRevB.102.235154} {\bibfield
  {journal} {\bibinfo  {journal} {Phys. Rev. B}\ }\textbf {\bibinfo {volume}
  {102}},\ \bibinfo {pages} {235154} (\bibinfo {year} {2020})}\BibitemShut
  {NoStop}%
\bibitem [{\citenamefont {Aditya}\ \emph {et~al.}(2022)\citenamefont {Aditya},
  \citenamefont {Samanta}, \citenamefont {Sen}, \citenamefont {Sengupta},\ and\
  \citenamefont {Sen}}]{Aditya2022prb}%
  \BibitemOpen
  \bibfield  {author} {\bibinfo {author} {\bibfnamefont {S.}~\bibnamefont
  {Aditya}}, \bibinfo {author} {\bibfnamefont {S.}~\bibnamefont {Samanta}},
  \bibinfo {author} {\bibfnamefont {A.}~\bibnamefont {Sen}}, \bibinfo {author}
  {\bibfnamefont {K.}~\bibnamefont {Sengupta}}, \ and\ \bibinfo {author}
  {\bibfnamefont {D.}~\bibnamefont {Sen}},\ }\href {\doibase
  10.1103/PhysRevB.105.104303} {\bibfield  {journal} {\bibinfo  {journal}
  {Phys. Rev. B}\ }\textbf {\bibinfo {volume} {105}},\ \bibinfo {pages}
  {104303} (\bibinfo {year} {2022})}\BibitemShut {NoStop}%
\bibitem [{\citenamefont {Makki}\ \emph {et~al.}(2022)\citenamefont {Makki},
  \citenamefont {Bandyopadhyay}, \citenamefont {Maity},\ and\ \citenamefont
  {Dutta}}]{Makki2022prb}%
  \BibitemOpen
  \bibfield  {author} {\bibinfo {author} {\bibfnamefont {A.~A.}\ \bibnamefont
  {Makki}}, \bibinfo {author} {\bibfnamefont {S.}~\bibnamefont
  {Bandyopadhyay}}, \bibinfo {author} {\bibfnamefont {S.}~\bibnamefont
  {Maity}}, \ and\ \bibinfo {author} {\bibfnamefont {A.}~\bibnamefont
  {Dutta}},\ }\href {\doibase 10.1103/PhysRevB.105.054301} {\bibfield
  {journal} {\bibinfo  {journal} {Phys. Rev. B}\ }\textbf {\bibinfo {volume}
  {105}},\ \bibinfo {pages} {054301} (\bibinfo {year} {2022})}\BibitemShut
  {NoStop}%
\bibitem [{\citenamefont {Zou}\ and\ \citenamefont {Ding}(2023)}]{Zou2023prb}%
  \BibitemOpen
  \bibfield  {author} {\bibinfo {author} {\bibfnamefont {Y.-T.}\ \bibnamefont
  {Zou}}\ and\ \bibinfo {author} {\bibfnamefont {C.}~\bibnamefont {Ding}},\
  }\href {\doibase 10.1103/PhysRevB.108.014303} {\bibfield  {journal} {\bibinfo
   {journal} {Phys. Rev. B}\ }\textbf {\bibinfo {volume} {108}},\ \bibinfo
  {pages} {014303} (\bibinfo {year} {2023})}\BibitemShut {NoStop}%
\bibitem [{\citenamefont {Cao}\ \emph {et~al.}(2024{\natexlab{a}})\citenamefont
  {Cao}, \citenamefont {Hu}, \citenamefont {Tong},\ and\ \citenamefont
  {Yang}}]{Cao2024prb109}%
  \BibitemOpen
  \bibfield  {author} {\bibinfo {author} {\bibfnamefont {K.}~\bibnamefont
  {Cao}}, \bibinfo {author} {\bibfnamefont {Y.}~\bibnamefont {Hu}}, \bibinfo
  {author} {\bibfnamefont {P.}~\bibnamefont {Tong}}, \ and\ \bibinfo {author}
  {\bibfnamefont {G.}~\bibnamefont {Yang}},\ }\href {\doibase
  10.1103/PhysRevB.109.024303} {\bibfield  {journal} {\bibinfo  {journal}
  {Phys. Rev. B}\ }\textbf {\bibinfo {volume} {109}},\ \bibinfo {pages}
  {024303} (\bibinfo {year} {2024}{\natexlab{a}})}\BibitemShut {NoStop}%
\bibitem [{\citenamefont {Lieb}\ \emph {et~al.}(1961)\citenamefont {Lieb},
  \citenamefont {Schultz},\ and\ \citenamefont {Mattis}}]{Lieb1961407}%
  \BibitemOpen
  \bibfield  {author} {\bibinfo {author} {\bibfnamefont {E.}~\bibnamefont
  {Lieb}}, \bibinfo {author} {\bibfnamefont {T.}~\bibnamefont {Schultz}}, \
  and\ \bibinfo {author} {\bibfnamefont {D.}~\bibnamefont {Mattis}},\ }\href
  {\doibase https://doi.org/10.1016/0003-4916(61)90115-4} {\bibfield  {journal}
  {\bibinfo  {journal} {Annals of Physics}\ }\textbf {\bibinfo {volume} {16}},\
  \bibinfo {pages} {407} (\bibinfo {year} {1961})}\BibitemShut {NoStop}%
\bibitem [{\citenamefont {Barouch}\ and\ \citenamefont
  {McCoy}(1971)}]{Barouch1971pra}%
  \BibitemOpen
  \bibfield  {author} {\bibinfo {author} {\bibfnamefont {E.}~\bibnamefont
  {Barouch}}\ and\ \bibinfo {author} {\bibfnamefont {B.~M.}\ \bibnamefont
  {McCoy}},\ }\href {\doibase 10.1103/PhysRevA.3.786} {\bibfield  {journal}
  {\bibinfo  {journal} {Phys. Rev. A}\ }\textbf {\bibinfo {volume} {3}},\
  \bibinfo {pages} {786} (\bibinfo {year} {1971})}\BibitemShut {NoStop}%
\bibitem [{\citenamefont {Bunder}\ and\ \citenamefont
  {McKenzie}(1999)}]{Bunder1999prb}%
  \BibitemOpen
  \bibfield  {author} {\bibinfo {author} {\bibfnamefont {J.~E.}\ \bibnamefont
  {Bunder}}\ and\ \bibinfo {author} {\bibfnamefont {R.~H.}\ \bibnamefont
  {McKenzie}},\ }\href {\doibase 10.1103/PhysRevB.60.344} {\bibfield  {journal}
  {\bibinfo  {journal} {Phys. Rev. B}\ }\textbf {\bibinfo {volume} {60}},\
  \bibinfo {pages} {344} (\bibinfo {year} {1999})}\BibitemShut {NoStop}%
\bibitem [{\citenamefont {Vitanov}\ and\ \citenamefont
  {Garraway}(1996)}]{Vitanov1996pra}%
  \BibitemOpen
  \bibfield  {author} {\bibinfo {author} {\bibfnamefont {N.~V.}\ \bibnamefont
  {Vitanov}}\ and\ \bibinfo {author} {\bibfnamefont {B.~M.}\ \bibnamefont
  {Garraway}},\ }\href {\doibase 10.1103/PhysRevA.53.4288} {\bibfield
  {journal} {\bibinfo  {journal} {Phys. Rev. A}\ }\textbf {\bibinfo {volume}
  {53}},\ \bibinfo {pages} {4288} (\bibinfo {year} {1996})}\BibitemShut
  {NoStop}%
\bibitem [{\citenamefont {Vitanov}(1999)}]{Vitanov1999pra}%
  \BibitemOpen
  \bibfield  {author} {\bibinfo {author} {\bibfnamefont {N.~V.}\ \bibnamefont
  {Vitanov}},\ }\href {\doibase 10.1103/PhysRevA.59.988} {\bibfield  {journal}
  {\bibinfo  {journal} {Phys. Rev. A}\ }\textbf {\bibinfo {volume} {59}},\
  \bibinfo {pages} {988} (\bibinfo {year} {1999})}\BibitemShut {NoStop}%
\bibitem [{\citenamefont {Zamani}\ \emph {et~al.}(2023)\citenamefont {Zamani},
  \citenamefont {Naji}, \citenamefont {Jafari},\ and\ \citenamefont
  {Langari}}]{zamani2310.15101}%
  \BibitemOpen
  \bibfield  {author} {\bibinfo {author} {\bibfnamefont {S.}~\bibnamefont
  {Zamani}}, \bibinfo {author} {\bibfnamefont {J.}~\bibnamefont {Naji}},
  \bibinfo {author} {\bibfnamefont {R.}~\bibnamefont {Jafari}}, \ and\ \bibinfo
  {author} {\bibfnamefont {A.}~\bibnamefont {Langari}},\ }\href@noop {}
  {\enquote {\bibinfo {title} {Scaling and universality at ramped quench
  dynamical quantum phase transition},}\ } (\bibinfo {year} {2023}),\ \Eprint
  {http://arxiv.org/abs/2310.15101} {arXiv:2310.15101 [cond-mat.stat-mech]}
  \BibitemShut {NoStop}%
\bibitem [{\citenamefont {Kou}\ and\ \citenamefont {Li}(2023)}]{Kou2023prb}%
  \BibitemOpen
  \bibfield  {author} {\bibinfo {author} {\bibfnamefont {H.-C.}\ \bibnamefont
  {Kou}}\ and\ \bibinfo {author} {\bibfnamefont {P.}~\bibnamefont {Li}},\
  }\href {\doibase 10.1103/PhysRevB.108.214307} {\bibfield  {journal} {\bibinfo
   {journal} {Phys. Rev. B}\ }\textbf {\bibinfo {volume} {108}},\ \bibinfo
  {pages} {214307} (\bibinfo {year} {2023})}\BibitemShut {NoStop}%
\bibitem [{\citenamefont {Fläschner}\ \emph {et~al.}(2018)\citenamefont
  {Fläschner}, \citenamefont {Vogel}, \citenamefont {Tarnowski}, \citenamefont
  {Rem}, \citenamefont {Lühmann}, \citenamefont {Heyl}, \citenamefont
  {Budich}, \citenamefont {Mathey}, \citenamefont {Sengstock},\ and\
  \citenamefont {Weitenberg}}]{Flaschner2018natphys14}%
  \BibitemOpen
  \bibfield  {author} {\bibinfo {author} {\bibfnamefont {N.}~\bibnamefont
  {Fläschner}}, \bibinfo {author} {\bibfnamefont {D.}~\bibnamefont {Vogel}},
  \bibinfo {author} {\bibfnamefont {M.}~\bibnamefont {Tarnowski}}, \bibinfo
  {author} {\bibfnamefont {B.~S.}\ \bibnamefont {Rem}}, \bibinfo {author}
  {\bibfnamefont {D.~S.}\ \bibnamefont {Lühmann}}, \bibinfo {author}
  {\bibfnamefont {M.}~\bibnamefont {Heyl}}, \bibinfo {author} {\bibfnamefont
  {J.~C.}\ \bibnamefont {Budich}}, \bibinfo {author} {\bibfnamefont
  {L.}~\bibnamefont {Mathey}}, \bibinfo {author} {\bibfnamefont
  {K.}~\bibnamefont {Sengstock}}, \ and\ \bibinfo {author} {\bibfnamefont
  {C.}~\bibnamefont {Weitenberg}},\ }\href {\doibase 10.1038/s41567-017-0013-8}
  {\bibfield  {journal} {\bibinfo  {journal} {Nature Physics}\ }\textbf
  {\bibinfo {volume} {14}},\ \bibinfo {pages} {265} (\bibinfo {year}
  {2018})}\BibitemShut {NoStop}%
\bibitem [{\citenamefont {Yu}(2017)}]{Yu2017pra96}%
  \BibitemOpen
  \bibfield  {author} {\bibinfo {author} {\bibfnamefont {J.}~\bibnamefont
  {Yu}},\ }\href {\doibase 10.1103/PhysRevA.96.023601} {\bibfield  {journal}
  {\bibinfo  {journal} {Phys. Rev. A}\ }\textbf {\bibinfo {volume} {96}},\
  \bibinfo {pages} {023601} (\bibinfo {year} {2017})}\BibitemShut {NoStop}%
\bibitem [{\citenamefont {Qiu}\ \emph {et~al.}(2018)\citenamefont {Qiu},
  \citenamefont {Deng}, \citenamefont {Guo},\ and\ \citenamefont
  {Yi}}]{Qiu2018pra98}%
  \BibitemOpen
  \bibfield  {author} {\bibinfo {author} {\bibfnamefont {X.}~\bibnamefont
  {Qiu}}, \bibinfo {author} {\bibfnamefont {T.-S.}\ \bibnamefont {Deng}},
  \bibinfo {author} {\bibfnamefont {G.-C.}\ \bibnamefont {Guo}}, \ and\
  \bibinfo {author} {\bibfnamefont {W.}~\bibnamefont {Yi}},\ }\href {\doibase
  10.1103/PhysRevA.98.021601} {\bibfield  {journal} {\bibinfo  {journal} {Phys.
  Rev. A}\ }\textbf {\bibinfo {volume} {98}},\ \bibinfo {pages} {021601}
  (\bibinfo {year} {2018})}\BibitemShut {NoStop}%
\bibitem [{\citenamefont {Lahiri}\ and\ \citenamefont
  {Bera}(2019{\natexlab{b}})}]{Lahiri2019prb99}%
  \BibitemOpen
  \bibfield  {author} {\bibinfo {author} {\bibfnamefont {A.}~\bibnamefont
  {Lahiri}}\ and\ \bibinfo {author} {\bibfnamefont {S.}~\bibnamefont {Bera}},\
  }\href {\doibase 10.1103/PhysRevB.99.174311} {\bibfield  {journal} {\bibinfo
  {journal} {Phys. Rev. B}\ }\textbf {\bibinfo {volume} {99}},\ \bibinfo
  {pages} {174311} (\bibinfo {year} {2019}{\natexlab{b}})}\BibitemShut
  {NoStop}%
\bibitem [{\citenamefont {Budich}\ and\ \citenamefont
  {Heyl}(2016)}]{Budich2016prb93}%
  \BibitemOpen
  \bibfield  {author} {\bibinfo {author} {\bibfnamefont {J.~C.}\ \bibnamefont
  {Budich}}\ and\ \bibinfo {author} {\bibfnamefont {M.}~\bibnamefont {Heyl}},\
  }\href {\doibase 10.1103/PhysRevB.93.085416} {\bibfield  {journal} {\bibinfo
  {journal} {Phys. Rev. B}\ }\textbf {\bibinfo {volume} {93}},\ \bibinfo
  {pages} {085416} (\bibinfo {year} {2016})}\BibitemShut {NoStop}%
\bibitem [{\citenamefont {Bhattacharya}\ and\ \citenamefont
  {Dutta}(2017)}]{Bhattacharya2017prb96}%
  \BibitemOpen
  \bibfield  {author} {\bibinfo {author} {\bibfnamefont {U.}~\bibnamefont
  {Bhattacharya}}\ and\ \bibinfo {author} {\bibfnamefont {A.}~\bibnamefont
  {Dutta}},\ }\href {\doibase 10.1103/PhysRevB.96.014302} {\bibfield  {journal}
  {\bibinfo  {journal} {Phys. Rev. B}\ }\textbf {\bibinfo {volume} {96}},\
  \bibinfo {pages} {014302} (\bibinfo {year} {2017})}\BibitemShut {NoStop}%
\bibitem [{\citenamefont {Bhattacharjee}\ and\ \citenamefont
  {Dutta}(2018)}]{Bhattacharjee2018prb97}%
  \BibitemOpen
  \bibfield  {author} {\bibinfo {author} {\bibfnamefont {S.}~\bibnamefont
  {Bhattacharjee}}\ and\ \bibinfo {author} {\bibfnamefont {A.}~\bibnamefont
  {Dutta}},\ }\href {\doibase 10.1103/PhysRevB.97.134306} {\bibfield  {journal}
  {\bibinfo  {journal} {Phys. Rev. B}\ }\textbf {\bibinfo {volume} {97}},\
  \bibinfo {pages} {134306} (\bibinfo {year} {2018})}\BibitemShut {NoStop}%
\bibitem [{\citenamefont {Zhou}\ \emph {et~al.}(2018)\citenamefont {Zhou},
  \citenamefont {Wang}, \citenamefont {Wang},\ and\ \citenamefont
  {Gong}}]{zhou2018pra98}%
  \BibitemOpen
  \bibfield  {author} {\bibinfo {author} {\bibfnamefont {L.}~\bibnamefont
  {Zhou}}, \bibinfo {author} {\bibfnamefont {Q.-h.}\ \bibnamefont {Wang}},
  \bibinfo {author} {\bibfnamefont {H.}~\bibnamefont {Wang}}, \ and\ \bibinfo
  {author} {\bibfnamefont {J.}~\bibnamefont {Gong}},\ }\href {\doibase
  10.1103/PhysRevA.98.022129} {\bibfield  {journal} {\bibinfo  {journal} {Phys.
  Rev. A}\ }\textbf {\bibinfo {volume} {98}},\ \bibinfo {pages} {022129}
  (\bibinfo {year} {2018})}\BibitemShut {NoStop}%
\bibitem [{\citenamefont {Lang}\ \emph
  {et~al.}(2018{\natexlab{b}})\citenamefont {Lang}, \citenamefont {Chen},
  \citenamefont {Hong},\ and\ \citenamefont {Fan}}]{Lang2018prb98}%
  \BibitemOpen
  \bibfield  {author} {\bibinfo {author} {\bibfnamefont {H.}~\bibnamefont
  {Lang}}, \bibinfo {author} {\bibfnamefont {Y.}~\bibnamefont {Chen}}, \bibinfo
  {author} {\bibfnamefont {Q.}~\bibnamefont {Hong}}, \ and\ \bibinfo {author}
  {\bibfnamefont {H.}~\bibnamefont {Fan}},\ }\href {\doibase
  10.1103/PhysRevB.98.134310} {\bibfield  {journal} {\bibinfo  {journal} {Phys.
  Rev. B}\ }\textbf {\bibinfo {volume} {98}},\ \bibinfo {pages} {134310}
  (\bibinfo {year} {2018}{\natexlab{b}})}\BibitemShut {NoStop}%
\bibitem [{\citenamefont {Yang}\ \emph {et~al.}(2018)\citenamefont {Yang},
  \citenamefont {Li},\ and\ \citenamefont {Chen}}]{Yang2018prb97}%
  \BibitemOpen
  \bibfield  {author} {\bibinfo {author} {\bibfnamefont {C.}~\bibnamefont
  {Yang}}, \bibinfo {author} {\bibfnamefont {L.}~\bibnamefont {Li}}, \ and\
  \bibinfo {author} {\bibfnamefont {S.}~\bibnamefont {Chen}},\ }\href {\doibase
  10.1103/PhysRevB.97.060304} {\bibfield  {journal} {\bibinfo  {journal} {Phys.
  Rev. B}\ }\textbf {\bibinfo {volume} {97}},\ \bibinfo {pages} {060304}
  (\bibinfo {year} {2018})}\BibitemShut {NoStop}%
\bibitem [{\citenamefont {Jafari}\ and\ \citenamefont
  {Akbari}(2021)}]{Jafari2021pra103}%
  \BibitemOpen
  \bibfield  {author} {\bibinfo {author} {\bibfnamefont {R.}~\bibnamefont
  {Jafari}}\ and\ \bibinfo {author} {\bibfnamefont {A.}~\bibnamefont
  {Akbari}},\ }\href {\doibase 10.1103/PhysRevA.103.012204} {\bibfield
  {journal} {\bibinfo  {journal} {Phys. Rev. A}\ }\textbf {\bibinfo {volume}
  {103}},\ \bibinfo {pages} {012204} (\bibinfo {year} {2021})}\BibitemShut
  {NoStop}%
\bibitem [{\citenamefont {Cao}\ \emph {et~al.}(2023{\natexlab{b}})\citenamefont
  {Cao}, \citenamefont {Yang}, \citenamefont {Hu},\ and\ \citenamefont
  {Yang}}]{cao2023prb108}%
  \BibitemOpen
  \bibfield  {author} {\bibinfo {author} {\bibfnamefont {K.}~\bibnamefont
  {Cao}}, \bibinfo {author} {\bibfnamefont {S.}~\bibnamefont {Yang}}, \bibinfo
  {author} {\bibfnamefont {Y.}~\bibnamefont {Hu}}, \ and\ \bibinfo {author}
  {\bibfnamefont {G.}~\bibnamefont {Yang}},\ }\href {\doibase
  10.1103/PhysRevB.108.024201} {\bibfield  {journal} {\bibinfo  {journal}
  {Phys. Rev. B}\ }\textbf {\bibinfo {volume} {108}},\ \bibinfo {pages}
  {024201} (\bibinfo {year} {2023}{\natexlab{b}})}\BibitemShut {NoStop}%
\bibitem [{\citenamefont {Cao}\ \emph {et~al.}(2024{\natexlab{b}})\citenamefont
  {Cao}, \citenamefont {Guo},\ and\ \citenamefont {Yang}}]{Cao2024jpcm36}%
  \BibitemOpen
  \bibfield  {author} {\bibinfo {author} {\bibfnamefont {K.}~\bibnamefont
  {Cao}}, \bibinfo {author} {\bibfnamefont {H.}~\bibnamefont {Guo}}, \ and\
  \bibinfo {author} {\bibfnamefont {G.}~\bibnamefont {Yang}},\ }\href {\doibase
  10.1088/1361-648X/ad1a5a} {\bibfield  {journal} {\bibinfo  {journal} {Journal
  of Physics: Condensed Matter}\ }\textbf {\bibinfo {volume} {36}},\ \bibinfo
  {pages} {155401} (\bibinfo {year} {2024}{\natexlab{b}})}\BibitemShut
  {NoStop}%
\bibitem [{\citenamefont {Yuzbashyan}\ \emph {et~al.}(2006)\citenamefont
  {Yuzbashyan}, \citenamefont {Tsyplyatyev},\ and\ \citenamefont
  {Altshuler}}]{Yuzbashyan2006prl}%
  \BibitemOpen
  \bibfield  {author} {\bibinfo {author} {\bibfnamefont {E.~A.}\ \bibnamefont
  {Yuzbashyan}}, \bibinfo {author} {\bibfnamefont {O.}~\bibnamefont
  {Tsyplyatyev}}, \ and\ \bibinfo {author} {\bibfnamefont {B.~L.}\ \bibnamefont
  {Altshuler}},\ }\href {\doibase 10.1103/PhysRevLett.96.097005} {\bibfield
  {journal} {\bibinfo  {journal} {Phys. Rev. Lett.}\ }\textbf {\bibinfo
  {volume} {96}},\ \bibinfo {pages} {097005} (\bibinfo {year}
  {2006})}\BibitemShut {NoStop}%
\bibitem [{\citenamefont {Sciolla}\ and\ \citenamefont
  {Biroli}(2010)}]{Sciolla2010prl}%
  \BibitemOpen
  \bibfield  {author} {\bibinfo {author} {\bibfnamefont {B.}~\bibnamefont
  {Sciolla}}\ and\ \bibinfo {author} {\bibfnamefont {G.}~\bibnamefont
  {Biroli}},\ }\href {\doibase 10.1103/PhysRevLett.105.220401} {\bibfield
  {journal} {\bibinfo  {journal} {Phys. Rev. Lett.}\ }\textbf {\bibinfo
  {volume} {105}},\ \bibinfo {pages} {220401} (\bibinfo {year}
  {2010})}\BibitemShut {NoStop}%
\bibitem [{\citenamefont {Ramos}\ \emph {et~al.}(2023)\citenamefont {Ramos},
  \citenamefont {Urichuk}, \citenamefont {Schneider},\ and\ \citenamefont
  {Sirker}}]{Ramos2023prb}%
  \BibitemOpen
  \bibfield  {author} {\bibinfo {author} {\bibfnamefont {F.~B.}\ \bibnamefont
  {Ramos}}, \bibinfo {author} {\bibfnamefont {A.}~\bibnamefont {Urichuk}},
  \bibinfo {author} {\bibfnamefont {I.}~\bibnamefont {Schneider}}, \ and\
  \bibinfo {author} {\bibfnamefont {J.}~\bibnamefont {Sirker}},\ }\href
  {\doibase 10.1103/PhysRevB.107.075138} {\bibfield  {journal} {\bibinfo
  {journal} {Phys. Rev. B}\ }\textbf {\bibinfo {volume} {107}},\ \bibinfo
  {pages} {075138} (\bibinfo {year} {2023})}\BibitemShut {NoStop}%
\bibitem [{\citenamefont {Cao}\ \emph {et~al.}(2024{\natexlab{c}})\citenamefont
  {Cao}, \citenamefont {Hu}, \citenamefont {Tong}, \citenamefont {Yang},\ and\
  \citenamefont {Liu}}]{cao2024relaxation}%
  \BibitemOpen
  \bibfield  {author} {\bibinfo {author} {\bibfnamefont {K.}~\bibnamefont
  {Cao}}, \bibinfo {author} {\bibfnamefont {Y.}~\bibnamefont {Hu}}, \bibinfo
  {author} {\bibfnamefont {P.}~\bibnamefont {Tong}}, \bibinfo {author}
  {\bibfnamefont {G.}~\bibnamefont {Yang}}, \ and\ \bibinfo {author}
  {\bibfnamefont {P.}~\bibnamefont {Liu}},\ }\href@noop {} {\enquote {\bibinfo
  {title} {Relaxation dynamics in the alternating xy chain following a quantum
  quench},}\ } (\bibinfo {year} {2024}{\natexlab{c}}),\ \Eprint
  {http://arxiv.org/abs/2311.08025} {arXiv:2311.08025 [cond-mat.stat-mech]}
  \BibitemShut {NoStop}%
\bibitem [{\citenamefont {Olver}\ \emph {et~al.}(2010)\citenamefont {Olver},
  \citenamefont {Lozier}, \citenamefont {Boisvert},\ and\ \citenamefont
  {Clark}}]{Olver2010NISTHO}%
  \BibitemOpen
  \bibfield  {author} {\bibinfo {author} {\bibfnamefont {F.~W.~J.}\
  \bibnamefont {Olver}}, \bibinfo {author} {\bibfnamefont {D.~W.}\ \bibnamefont
  {Lozier}}, \bibinfo {author} {\bibfnamefont {R.~F.}\ \bibnamefont
  {Boisvert}}, \ and\ \bibinfo {author} {\bibfnamefont {C.~W.}\ \bibnamefont
  {Clark}}\ }(\bibinfo {year} {2010})\BibitemShut {NoStop}%
\end{thebibliography}%

\end{document}